\documentclass[fleqn,usenatbib]{mnras}
\usepackage{newtxtext,newtxmath}
\usepackage[T1]{fontenc}
\usepackage{ae,aecompl}

\usepackage{graphicx}	% Including figure files
\usepackage{amsmath}	% Advanced maths commands
\usepackage{amssymb}	% Extra maths symbols
\usepackage{enumerate}
\title[Constraining radio flares from compact binary mergers]{Constraining coherent low frequency radio flares from compact binary mergers}

\author[A. Rowlinson \& G.E. Anderson.]{
A. Rowlinson$^{1,2}$\thanks{E-mail: b.a.rowlinson@uva.nl}, G.E. Anderson$^{3}$ 
\\
$^{1}$ Anton Pannekoek Institute, University of Amsterdam, Postbus 94249, 1090 GE, Amsterdam, The Netherlands\\
$^{2}$ Netherlands Institute for Radio Astronomy (ASTRON), PO Box 2, 7990 AA Dwingeloo, The Netherlands \\
$^{3}$ International Centre for Radio Astronomy Research, Curtin University, GPO Box U1987, Perth, WA 6845, Australia
}

\date{Accepted XXX. Received YYY; in original form ZZZ}

\pubyear{2015}

\begin{document}
% * <antoniarowlinson@gmail.com> 2017-08-02T13:25:12.082Z:
%
% ^.
\label{firstpage}
\pagerange{\pageref{firstpage}--\pageref{lastpage}}
\maketitle

\begin{abstract}
The presence and detectability of coherent radio emission from compact binary mergers (containing at least one neutron star) remains poorly constrained due to large uncertainties in the models. These compact binary mergers may initially be detected as Short Gamma-ray Bursts (SGRBs) or via their gravitational wave emission. Several radio facilities have developed rapid response modes enabling them to trigger on these events and search for this emission. For this paper, we constrain this coherent radio emission using the deepest available constraints for GRB 150424A, which were obtained via a triggered observation with the Murchison Widefield Array. We then expand this analysis to determine the properties of magnetar merger remnants that may be formed via a general population of binary neutron star mergers. Our results demonstrate that many of the potential coherent emission mechanisms that have been proposed for such events can be detected or very tightly constrained by the complementary strategies used by the current generation of low-frequency radio telescopes. 
\end{abstract}


%%%%%%%%%%%%%%%%%%%%%%%%%%%%%%%%%%%%%%%%%%%%%%%%%%

%%%%%%%%%%%%%%%%% BODY OF PAPER %%%%%%%%%%%%%%%%%%

\section{Introduction}

On 17th August 2017, the inspiral and merger of two neutron stars was detected by the Advanced Laser Interferometer Gravitational-Wave Observatory and the Advanced Virgo gravitational-wave observatory \citep[aLIGO and aVirgo;][]{abbott2017a}. In a momentous development for the field of multi-messenger astronomy, the electromagnetic counterpart was also rapidly identified, located and studied at a wide range of wavelengths and over a wide range of timescales \citep{abbott2017b}. This wealth of information is leading to a number of open questions and excitement for the discoveries that the future may hold in this field in the coming years.

A number of theories have predicted early time, low frequency, coherent radio emission from compact binary mergers via a wide range of emission mechanisms \citep[][]{usov2000, moortgat2003, pshirkov2010,  totani2013, falcke2014, zhang2014, mingarelli2015, wang2016}. This emission can occur in the course of the final seconds prior to the merger, during the merger or produced by the remnant following the merger. Detection of coherent radio emission associated with binary mergers, and the timescale it is observed on, can provide clues about the physical mechanisms involved, the nature of the remnant and the equation of state of nuclear matter \citep[e.g.][]{lasky2014}. This provides an intriguing potential electromagnetic counterpart to aLIGO triggers \citep{chu2016} and an interesting contender for the progenitor of some of the population of fast radio bursts \citep[FRBs;][]{zhang2014}.

The first searches for prompt coherent radio emission associated with compact binary mergers were performed by targeting the known population of SGRBs, which are expected to share the same progenitor. However, these searches were typically insufficiently sensitive \citep[e.g.][]{balsano1998} or potentially triggered too late following the SGRB for the given observing frequency \citep[][]{bannister2012}. 

The wide fields of view and rapid respointing capabilities of the next generation of low frequency radio telescopes, such as the Murchison Widefield Array \citep[MWA;][]{tingay2013}, Long Wavelength Array \citep[LWA;][]{taylor2012,ellingson2013} and the LOw Frequency ARray \citep[LOFAR;][]{haarlem2013}, will allow us to detect or tightly constrain predicted coherent radio signals. 
In fact, the first deep observational limits on coherent radio emission from compact mergers were obtained for GRB 150424A using the MWA \citep{kaplan2015} and for GRB 170112A using the LWA \citep{anderson2017}. Both of these GRBs were detected using the Neil Gehrels {\it Swift} Observatory \citep[hereafter referred to as {\it Swift};][]{gehrels2004}. 

GRB 150424A had a bright X-ray afterglow, which shows evidence of on-going energy injection into the system by the merger remnant \citep{melandri2015,rowlinson2013}. These X-ray counterparts can give clues to the coherent radio emission expected from these systems, so this paper focuses on the constraints that can be made using GRB 150424A whereas GRB 170112A was undetected in X-rays \citep{dai2017}. GRB 150424A was a SGRB with extended emission \citep[EE SGRBs;][]{norris2006}; EE SGRBs are GRBs whose durations place them in the category of Long GRBs (LGRBs), but whose properties are more consistent with the SGRB population leading to the expectation that they share a common progenitor, namely a compact binary merger \citep[e.g.][]{lattimer1976, eichler1989}.  

In this Paper, we investigate the potential of detecting coherent radio emission from the merger of compact binary systems. In Section \ref{sec:coherent}, we consider the propagation effects that coherent radio emission will experience as it travels through the Universe and the theoretical models describing the potential origins of this coherent radio emission. We then examine the current and soon to be available radio telescope capabilities in the context of pursuing this emission in Section \ref{sec:telescopes}. Following on from this, we apply the theoretical models to the current prompt observations of GRBs (outlined in Section \ref{sec:application}) before describing our expectations for future observations (Section \ref{sec:future}).

Throughout this work, we adopt a cosmology with $H_0 = 71$ km s$^{−1}$ Mpc$^{−1}$,  $\Omega_m = 0.27$ and $\Omega_{\lambda}= 0.73$. Errors are quoted at 90 per cent confidence for X-ray data and at 1$\sigma$ for fits to the magnetar model.

\section{Models of coherent radio emission from neutron star binary mergers}
\label{sec:coherent}

\begin{figure*}
\centering
\includegraphics[width=0.9\textwidth]{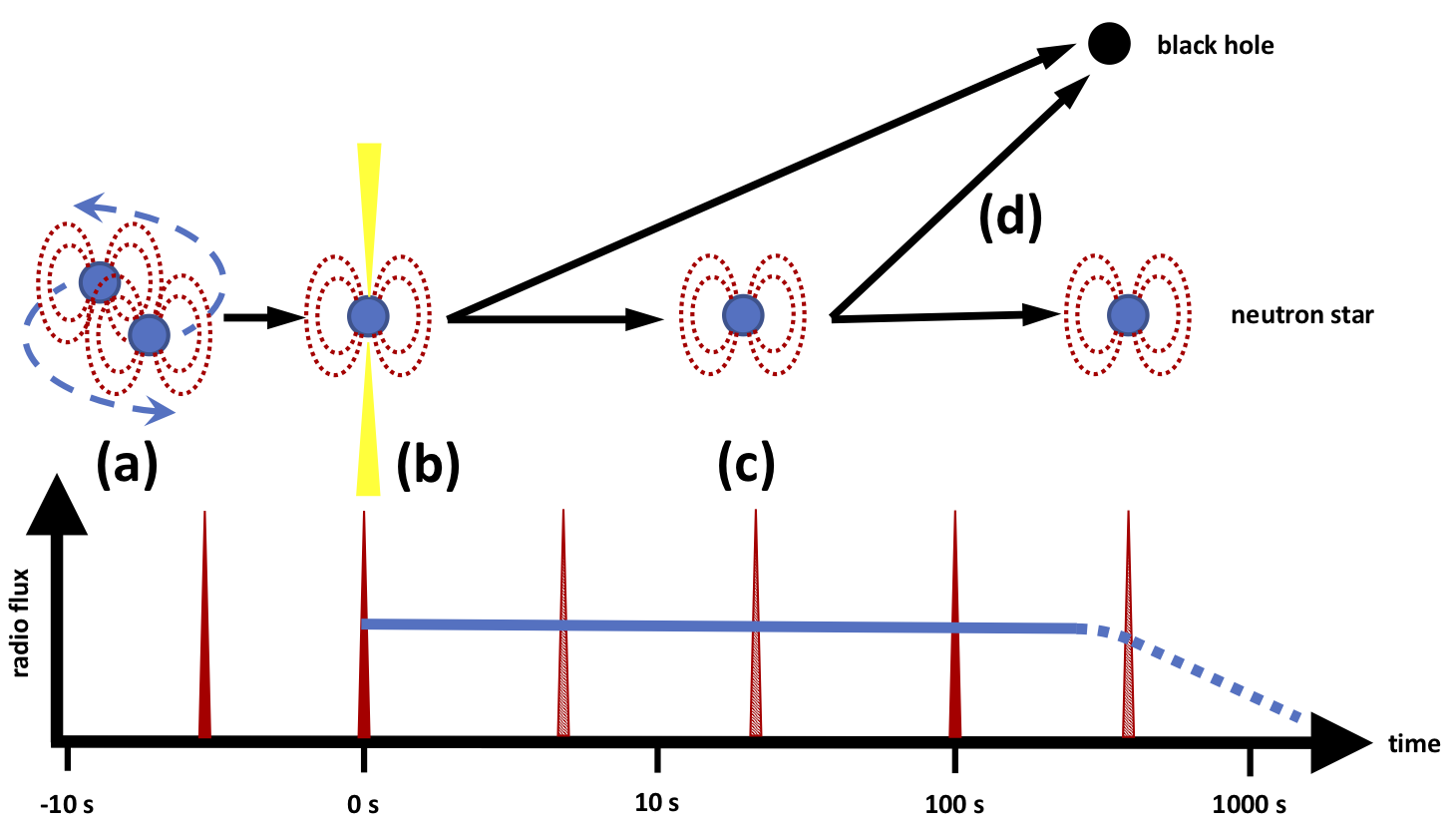}
\caption{
An illustration of the outcomes of a binary neutron star merger and the various coherent emission models as a function of the restframe timescale from the merger time. In the rough lightcurve, the red spikes are millisecond duration radio flares, consistent with fast radio bursts (FRBs), while the blue line represents persistent radio emission.  During the inspiral phase (a), the gravitational waves may excite the surrounding plasma causing short duration radio emission \citep[e.g.][]{moortgat2003} or brief flashes of radio emission can be caused by interactions between the magnetic fields of the neutron stars just prior to the merger \citep[e.g.][]{lipunov1996,metzger2016} (see Section \ref{sec:priorMerge}). When the two neutron stars merge, phase (b), they launch a highly relativistic jet that can produce a coherent burst of radio emission when interacting with the interstellar medium \citep[][see Section \ref{sec:jetInteract}]{usov2000}. At this point, the merger remnant will either collapse to form a black hole or a millisecond spin period neutron star with strong magnetic fields (magnetar). If a neutron star is formed, phase (c), there are a number of mechanisms to produce coherent radio emission. This emission is standard pulsar dipole radiation powered by the spin down of the neutron star, represented by the blue line, lasting for the lifetime of the neutron star that quickly fades with the rapid spin down of the neutron star \citep[e.g.][see Section \ref{sec:pulsar}]{totani2013}. Alternatively, the rapidly rotating neutron star may be able to produce repeating FRBs \citep[e.g. FRB 121102;][]{spitler2016}, represented by the red hatched spikes, during its lifetime while its spin and magnetic field remain sufficiently high \citep[e.g.][see Section \ref{sec:repeatFRB}]{metzger2017}. Finally, if the mass of the newly formed neutron star is too high, it will collapse to form a black hole producing a final FRB at the time of collapse due to the magnetic reconnection of the magnetosphere \citep[phase (d);][see Section \ref{sec:collapse}]{zhang2014} and all other coherent emission will immediately cease.}
\label{fig:CohModels}
\end{figure*}

A number of theories predict that coherent radio emission should arise from the merger of a compact binary system or from on-going activity of the magnetar central engine. In this section, we will first consider the propagation effects that coherent emission is subjected to, a description of the proposed central engines that could be created by compact binary mergers that power the emission, followed by an overview of the main theoretical models predicting coherent emission from these events. In Figure \ref{fig:CohModels}, we provide a cartoon diagram illustrating the timescales of each emission model relative to the evolution of a neutron star binary merger.

\subsection{Propagation effects}

\subsubsection{Dispersion}
Any coherent radio emission emitted by compact binary mergers will have been subjected to a dispersive delay caused by the intervening medium (the interstellar and intergalactic media) between us and the source. The dispersion delay, $\tau$, is given by
\begin{eqnarray}
\tau = \frac{\rm DM}{241.0 \nu^2_{\mathrm{GHz}}} ~{\rm s} \label{eqn:disp_delay}
\end{eqnarray}
where DM is the dispersion measure in pc cm$^{-3}$ and $\nu_{\mathrm{GHz}}$ is the observing frequency in GHz \citep{taylor1993}. Although the dispersion caused by the intergalactic medium is poorly known, we can estimate this value using
\begin{equation}
{\rm DM} \sim 1200 z ~{\rm pc ~cm}^{-3} \label{eqn:DM}
\end{equation}
where $z$ is the redshift of the source \citep[e.g.][]{ioka2003,inoue2004,lorimer2007,karastergiou2015}\footnote{See also \cite{dolag2015} for an alternative method to constrain the redshift.}.

\subsubsection{Plasma absorption}
When searching for coherent emission at low radio frequencies, one concern is that such signals will be absorbed below the plasma frequency, $\nu_p$, where the plasma becomes opaque. While the emitting region is unaffected by this, any plasma between the source and the observer will block the emission below this frequency \citep[for instance, solar flares where the observing frequencies can be used to probe different depths of the solar corona; e.g.][]{wild1963}. The plasma frequency is given by:
\begin{eqnarray}
\nu_{p} \simeq 9 n_{e} ~{\rm kHz} \label{eqn:plasFreq}
\end{eqnarray}
where $n_e$ is the number density of electrons in cm$^{-3}$. In the case of a relativistic plasma, with a Lorentz factor $\gamma$, this equation becomes:
\begin{eqnarray}
\nu_{p} \simeq 9 \left( \frac{n_{e}}{\gamma} \right) ~{\rm kHz} \label{eqn:plasFreqRel}
\end{eqnarray}

In the specific context of a GRB blast wave, \cite{zhang2014} shows that the co-moving plasma frequency is given by
\begin{eqnarray}
\nu_p \simeq 3.8\times10^{6} L_{52}^{\frac{1}{2}} \Gamma_{2}^{-1} r_{17}^{-1}~\mathrm{Hz} \label{GRB_plasmafreq}
\end{eqnarray}
where $r_{17}$ is the blast wave radius in units of $10^{17}$\,cm, $\Gamma = 100 \Gamma_2$ is the bulk Lorentz factor of the blast wave and $L_{52}$ is the gamma-ray luminosity of the GRB in units of $10^{52}$\,erg\,s$^{-1}$. 

The plasma frequency in the shocked material behind the GRB blast wave is expected to be much lower than the value obtained in the blast wave \citep{zhang2014}. Therefore, any coherent emission at $\nu_{\rm obs} \ge 0.12$ MHz is expected to pass through the relativistic blast wave.
However, as shown by \cite{zhang2014}, off-axis emission from the relativistic jet is likely to be blocked by the ejecta launched by the merger process, which goes on to form the kilonova. \cite{yamasaki2018} conduct a hydrodynamic simulation to investigate the propagation of coherent radio emission and find that the dynamical ejecta will hinder the escape of coherent radio emission after $\sim$1 millisecond until $\sim$1--10 years after the merger. The probability of detection will then depend on the isotropy of the ejecta. Simulations show the ejecta tend to be non-isotropic and typically close to the equatorial plane \citep[e.g.][]{rosswog1999}, hence there is still a chance that some emission will be able to escape for off-axis viewing angles. 

\subsubsection{Other factors}
In addition to the plasma frequency constraint, there are other factors that affect the emission as a function of the frequency. Interactions of the photons with the surrounding medium will lead to the thermal process of free-free absorption, resulting in a spectral break at low frequencies with a dependency of $\nu^{2}$. The frequency at which free-free absorption dominates is strongly dependent on the temperature and density of the surrounding medium. SGRBs are known to occur in very low density environments \citep{fong2015b} so although the temperature is poorly known, we anticipate that the free-free absorption is negligible for SGRBs at low radio frequencies. Additionally, for the population of FRBs observed at 1.4 GHz, it has recently been shown that free-free absorption is unlikely to play a significant role in their non-detection at MWA observing frequencies \citep{sokolowski2018}.

A further propagation effect that may impact the short duration pulses outlined in the following sections (though not the persistent emission model) is temporal smearing or scattering. Temporal smearing is the process by which short duration pulses are scattered by an inhomogenious plasma between the source and the observer, leading to a lengthening of the pulse duration, which causes the peak flux density to be reduced. In the same FRB study mentioned above, \cite{sokolowski2018} also demonstrated that temporal smearing had a negligible impact on the MWA observations.

\subsection{Review of central engine models}

Two central engine models have been proposed for GRBs: a black hole and a milli-second spin period, hyper-massive neutron star with extreme magnetic fields, which is commonly referred to as a `magnetar' and is the term we adopt for this paper. In this section, we first review the different models used to describe ongoing central engine activity observed in GRBs and then describe the magnetar central engine model in more depth as many of the coherent emission models depend upon the formation of a magnetar.

\subsubsection{Central engine activity inferred from X-ray light curves}
\label{sec:centralEng}

In the population of GRBs, and particularly the SGRBs associated with compact binary mergers, there is evidence of prolonged energy injection following the initial burst of gamma-rays. For SGRBs, the emission is typically separated as being ongoing gamma-ray/hard X-ray emission for $\sim$100 seconds following the GRB, also known as extended emission (e.g. Figure \ref{fig:XRTlc} phase (ii)), and a late-time X-ray plateau phase lasting from 100s of seconds to a few hours after the GRB (e.g. Figure \ref{fig:XRTlc} phase (iv)).

Different models have been proposed to explain the observed gamma-ray/hard X-ray extended emission:
\begin{itemize}
  \item Fallback accretion onto a black hole central engine \citep{rosswog2007}
  \item The launching of two jets by the engine \citep{barkov2011}
  \item r-process heating of the accretion disk \citep{metzger2010}
  \item Magnetic reconnection and turbulence \citep{zhang2011}
  \item Residual dipole spin-down energy of a magnetar  \citep{metzger2008, bucciantini2012}
  \item Magnetic propellering of material attempting to accrete onto a magnetar central engine \citep{gompertz2013,gompertz2014}
\end{itemize}

While the models proposed for the X-ray plateau phases include:
\begin{itemize}
  \item On-going accretion onto a black hole central engine \citep{cannizzo2011}
  \item Residual dipole spin-down energy of a magnetar \citep{zhang2001, rowlinson2013,gompertz2014}
\end{itemize}

In the the compact binary merger progenitor scenario, accretion onto the compact remnant is expected to be within the first 2 seconds \citep[][]{rezzolla2011}, with only a small amount of material on highly eccentric orbits that can accrete at late times to produce an X-ray flare \citep[e.g. ][]{rosswog2007}. It is therefore unlikely that on-going accretion can explain the observed energy injection so magnetar central engine models are typically considered to be the origin of this emission \citep{rowlinson2013}. 

\subsubsection{Magnetar central engine model}
\label{sec:centralEng2}

Assuming that the late-time X-ray emission (phases (ii) and (iv) in Figure \ref{fig:XRTlc}) observed from SGRBs is due to magnetar central engines, the model dependent properties of the magnetar can be used to constrain the coherent radio emission expected. By fitting rest-frame X-ray light curves, and utilising the method outlined in \cite{rowlinson2013}, we can constrain the magnetar central engine model parameters. Following the spin-down model of \cite{zhang2001}, the bolometric luminosity, $L = 10^{49} L_{0,49}$ erg s${-1}$, and duration of the X-ray plateaus, $T = 10^{3} T_{{\rm em}, 3}$ s, can be directly linked to the magnetic field, $B=10^{15} B_{15}$ G, and the initial spin period, $P = 10^{-3} P_{-3}$ s, of the magnetar using
\begin{eqnarray}
B_{15}^2 = 4.2025 M_{1.4}^2 R_{6}^{-2}L_{0,49}^{-1}T_{{\rm em}, 3}^{-2}\left( \frac{\epsilon}{1 - \cos \theta} \right), \label{eqn:B} \\
P_{-3}^{2} = 2.05 M_{1.4}R_{6}^{2}L_{0,49}^{-1}T_{{\rm em}, 3}^{-1}\left( \frac{\epsilon}{1 - \cos \theta} \right), \label{eqn:P}
\end{eqnarray}
where $R = 10^{6} R_6$ cm is the radius of the magnetar, $M = 1.4 M_{1.4}$ M$_{\odot}$ is the mass of the magnetar, $\epsilon$ is the efficiency of converting rotational energy into X-ray emission and $\theta$ is the jet opening angle in degrees. The magnetar will be rapidly spinning down via dipole radiation, leading to a decline in the observed luminosity, $L_{{\rm em},49}$, described by
\begin{eqnarray}
L_{{\rm em},49}(T) = L_{0,49} \left( 1 + \frac{T}{10^3T_{{\rm em},3}} \right)^{-2} \label{eqn:Lumin}
\end{eqnarray}
where $T$ is the time in seconds after formation of the magnetar. We assume the magnetar has a typical neutron star radius of $10^6$ cm and has a mass of 2.1 M$_{\odot}$ (consistent with formation via the merger of two typical neutron stars). Additionally, from the analysis of \cite{rowlinson2014}, we are able to use the observed correlation between the luminosity of X-ray plateaus and their durations \citep{dainotti2008, dainotti2010, dainotti2013} to place constraints on the factor $f = \left( \frac{\epsilon}{1 - \cos \theta} \right)$, which encompasses the uncertainty in the beaming angles and efficiencies. By using all the combinations of $\epsilon$ and $\theta$ within the 95\% probability contours, we estimate this ratio to be $f = 3.45\pm0.29$ for all magnetars formed during compact binary mergers. 

We note that the initial spin period may be higher than that obtained in both of these fits, as the magnetar is likely to have spun down due to gravitational wave losses \citep[e.g.][]{zhang2001,corsi2009} and/or via magnetic propellering \citep{gompertz2013, gompertz2014}. 

\subsection{Coherent radio emission prior to merger}
\label{sec:priorMerge}

The models described in this section are applicable to any neutron star binary system, as they are not dependent on the magnetar's spin period or magnetic field strength. Additionally, they are independent of the central engine formed. In Figure \ref{fig:CohModels}, these emission processes occur during phase (a).

\subsubsection{Alignment of magnetic fields}
\label{sec:allignB}

\cite{lipunov1996} proposed that during the final stages of the merger process, neutron stars can be spun up to millisecond spin periods. This leads to a revival of the pulsar emission mechanism in the binary and they propose four models by which energy can be extracted from the system via magnetic dipole or quadrupole radiation up to a few seconds prior to the merger timescale. They predict that the brightest emission will occur at the time of first contact of the neutron star surfaces, giving a burst of radio emission. This idea has been expanded further by \cite{hansen2001, lyutikov2013} and \cite{lyutikov2013} predict that the peak luminosity of this pulsar-like coherent emission, occurring at the merger time, is given by
\begin{eqnarray}
L_{\rm max} \sim 3\times 10^{50} \frac{B_{15}^2 M_{1.4}^3}{R_6} ~{\rm erg}~{\rm s}^{-1}. \label{eqn:lyutikov2013}
\end{eqnarray}
\cite{lyutikov2013} assume that a small fraction, $\epsilon_r$, of the wind luminosity is converted into coherent radio emission with a spectral slope of -1. The observable flux density, $F_{\nu}$, at the observed frequency ($\nu_{\rm obs} = (1+z) \nu_{\rm rest}$) is therefore predicted to be
\begin{eqnarray}
\begin{aligned}
F_{\nu} &\simeq \frac{L \epsilon_r (1+z)}{4\pi d^2 \nu_{\rm obs}} ~{\rm erg}~{\rm cm}^{-2}~{\rm s}^{-1}~{\rm Hz}^{-1}\\
& \sim 2\times10^{8} (1+z) \frac{B_{15}^2 M_{1.4}^3 }{R_6 \nu_{9, {\rm obs}} D^2} \epsilon_r ~{\rm Jy} \label{eqn:lyutikov2013_2}
\end{aligned}
\end{eqnarray}
where $D$ is the distance in Gpc (and $d$ is the distance to the source in meters), $\nu = 10^9 \nu_9$. with typical neutron star parameters of $M_{1.4}=1$, $R_6 = 1$ and $B_{15}=10^{-3}$. the typical value of $\epsilon_r = 10^{-4}$ for known pulsars \citep[e.g.][]{taylor1993b}.

\subsubsection{Other models}
\label{sec:othermodels}
In addition to the model considered in Section \ref{sec:allignB}, there are other models predicting early time coherent radio emission. These models are not considered further in this publication, but are briefly described here for completeness.

Firstly, during the final stages of a compact binary merger, there will be large amplitude gravitational waves propagating through the hot, magnetised plasma formed by the neutron stars as they get destroyed. \cite{moortgat2003,moortgat2004,moortgat2006} predict that this will excite low frequency magnetohydrodynamic waves that will produce observable, coherent radio emission. \cite{moortgat2004} postulate a binary containing a magnetar and show that the total energy released from gravitational waves via this mechanism would be $10^{50}$ erg.

Secondly, if there is at least one magnetised neutron star (with a typical magnetic field of $10^{12}$ G) in the binary system then during the final stages prior to the merger, its orbital motion around its companion will induce a strong current along the magnetic field lines as the two sources interact. The electric current will lead to observable synchrotron emission. This scenario has been considered in the case of a binary neutron star system \citep{metzger2016} and in a neutron star - black hole binary \citep{mingarelli2015}.

\cite{metzger2016} show that for a binary neutron star system, this emission will be optically thick and lead to a pair wind fireball whose emission will be quasi-thermal. This does not lead to a detectable coherent radio signal, however, they postulate that there could be magnetic reconnection and particle acceleration in the region outside of the pair photosphere that would lead to a signal similar to an FRB.

In the case of a black hole battery (a system containing one neutron star and a black hole), \cite{mingarelli2015} show that the potential energy release is comparable to that of an FRB. They also suggest that there may be a second peak in the luminosity when the neutron star is fully accreted onto the black hole and its magnetic field undergoes reconnection \citep[similar to that postulated by ][ and outlined in Section \ref{sec:collapse}]{falcke2014}.

{In these models,} outside of the photosphere the environment will be extremely low density, allowing any coherent radio emission to pass unimpeded.

\subsection{Coherent emission from interaction of relativistic jet with ISM}
\label{sec:jetInteract}

The composition of GRB jets remains relatively unconstrained, with two main models proposed: a baryonic jet \citep{paczynski1986, goodman1986} and a Poynting flux dominated wind \citep{usov1994, thompson1994}. \cite{usov2000} proposed that if powered by a Poynting flux dominated wind, the magnetic field at the shock front would be high. This high magnetic field, coupled with the high Lorentz factors, would lead to the generation of low frequency radio emission at the shock front (occurring during phase (b) in Figure \ref{fig:CohModels}). In order to generate the highly magnetised wind, they consider a fast rotating and highly magnetised central engine, such as the magnetars proposed as GRB central engines. \cite{usov2000} showed that the magnetised wind properties can be directly linked to the properties of the magnetar powering the GRB given by Equations \ref{eqn:B} and \ref{eqn:P}, where the strength of the magnetic field at the shock front (also known as the deceleration radius) in the rest frame of the wind, $B_0$, is given by
\begin{eqnarray}
%\begin{split}
B_0 & = \frac{B{\rm s}}{\Gamma_0} \\
& \simeq 18 B_{15} P_{-3}^{-2} \epsilon_{B}^{\frac{1}{2}}  \left( \frac{n}{Q_{53} \Gamma_{0}}\right) ^{\frac{1}{3}} ~{\rm G} \label{eqn:B0}
%\end{split}
\end{eqnarray}
where $B_{\rm s}$ is the magnetic field at the shock front, $n \sim 1$ cm$^{-3}$ is the density of the surrounding medium, $\epsilon_B$ is the fraction of the wind energy contained within the magnetic field at the deceleration radius, $\Gamma_{0}$ is the Lorentz factor of the relativistic wind and $Q = 10^{53} Q_{53}$ erg is the initial kinetic energy of the wind. As the coherent radio emission is associated with the interaction of the shock front producing the GRB, the coherent radio emission is expected to occur during the prompt gamma-ray emission phase. Additionally, as the emission region is at the shock front, the emission only needs to propagate through the low density surrounding medium leading to a very low plasma frequency (of order kHz) and hence emission at MHz frequencies should be detectable.

This magnetic field strength at the shock front in the rest frame of the wind can be used to determine the  peak frequency of the coherent radio emission:
\begin{eqnarray}
\nu_{\rm max} \simeq \frac{1}{1 + z} \frac{B_{0}}{100} ~{\rm MHz} \label{eqn:nuMax}
\end{eqnarray}

This emission, which relies on the relativistic wind producing the GRB, is expected to have an un-disperesed duration, $\tau_r$, which is less than the GRB.

We typically expect to be in the dispersion limited regime ( where the dispersion delay is longer than the duration of the coherent radio emission; $\frac{2\Delta \nu}{\nu}\tau \geq \tau_r$) with a radio spectral flux density given by:
\begin{eqnarray}
F_{\nu} \simeq \frac{0.1 \epsilon_{B} (\beta - 1)}{2 \Delta \nu \tau} \left( \frac{\nu_{\rm obs}}{\nu_{\rm max}} \right)^{1-\beta} \Phi_{\gamma} ~{\rm erg ~cm}^{-2} {\rm s}^{-1} {\rm Hz}^{-1} \label{eqn:UsovFlux1}
\end{eqnarray}
where $\beta \simeq 1.6$ is the spectral index of the emission, $\Delta \nu$ is the bandwidth of the observation in Hz, and $\Phi_{\gamma}$ is the observed fluence in gamma-rays (in erg cm$^{-2}$). Following \cite{trott2013}, we can estimate that the observed flux density, $F_{\nu,{\rm obs}}$, of a dispersed pulse in snapshot images (in which the snapshot integration time, $t_{\rm int}$ is longer than the pulse duration) is given by:
\begin{eqnarray}
F_{\nu,{\rm obs}} = F_{\nu} \left( \frac{\tau_r}{t_{\rm int}} \right) \label{eqn:obsFlx}
\end{eqnarray}
For longer duration pulses, the emission will not be dispersion limited  ($\frac{2\Delta \nu}{\nu}\tau < \tau_r$) and the flux density is given by:
\begin{eqnarray}
F_{\nu} \simeq \frac{0.1 \epsilon_{B} (\beta - 1)}{\tau_r \nu_{\rm max}} \left( \frac{\nu}{\nu_{\rm max}} \right)^{-\beta} \Phi_{\gamma} ~{\rm erg ~cm}^{-2} {\rm s}^{-1} {\rm Hz}^{-1}. \label{eqn:UsovFlux2}
\end{eqnarray}
In this case, the observed flux density is equal to the flux density of the un-dispersed pulse.

\subsection{Coherent radio emission from late time central engine activity}
\label{sec:latetime}

The late time central engine activity occurs during phases (c) and (d) as shown in Figure \ref{fig:CohModels}.

\subsubsection{Persistent, pulsar-like emission}
\label{sec:pulsar}

In the model proposed by \cite{totani2013} \citep[see also ][]{pshirkov2010}, during the final stages of the merger of two neutron stars the magnetic fields will align with the binary rotation axis. Once aligned, they propose the binary will emit via a similar mechanism to isolated pulsars and this emission will continue until the magnetar, thought to be formed by the merged neutron stars, collapses to form a black hole (n.b. not all of the newly formed magnetars will collapse). The emission is expected to occur via magnetic breaking, e.g. dipole spin-down, which will be powered by the magnetic field of the individual neutron stars prior to merger and the amplified magnetic field of the magnetar following the merger \citep[ignoring losses due to gravitational wave emission; e.g.][]{corsi2009}.

The pulsar-like emission may start shortly before the merger when the magnetic fields of the two neutron stars align (phase (a) in Figure \ref{fig:CohModels}). However, this emission is expected to be longer lived following the formation of the magnetar (phase (c) in Figure \ref{fig:CohModels}), only switching off if/when it collapses to form a black hole. Therefore, the timescale of this emission is dependent upon the mass of the magnetar formed, perhaps lasting a few seconds to a few hours or, in the case of the formation of a stable magnetar, it may be continuous but decreasing in luminosity as it spins down.

Due to the long emission timescales expected for this model, we can assume that we are in a non-dispersed regime, meaning that we do not need to correct the predicted fluxes to account for dispersion in snapshot images (see Equation \ref{eqn:obsFlx}). Assuming that the emission process is comparable to that of known pulsars, \cite{totani2013} predict the neutron star will have a flux density of:
\begin{eqnarray}
F_{\nu} \simeq 8\times10^{7} \nu_{\rm obs}^{-1} \epsilon_{r} D^{-2} 
B_{15}^{2} R_6^6 P_{-3}^{-4} ~{\rm Jy} \label{eqn:totaniFlx}
\end{eqnarray}
where $\nu_{\rm obs}$ is the frequency in MHz and $\epsilon_r$ is the efficiency.
The efficiency is the remaining unknown quantity in this analysis, with pulsars typically having a value of $10^{-4}$ \citep[e.g.][]{taylor1993b}.

As this emission is postulated to be pulsar-like, we can assume that it will be beamed in the same way as typical pulsars. Therefore, the emission is going to be along the magnetic axis of the magnetar. Due to the dynamo mechanism that enhanced the magnetic field strength, this axis is most likely to be aligned with the rotation axis of the magnetar \citep{cheng2014, giacomazzo2014} although it may become misaligned with time \citep{cutler2002}. The likely alignment of the pulsar beams with the rotation axis means that it will be along the same axis as the relativistic jet producing the GRB. For off-axis viewing angles, it may not be possible to see this emission due to the beaming of the pulsar emission. It is likely that the pulsar emission cone is wider at lower radio frequencies \citep[from the interpretation of pulse width observations; e.g.][]{mitra2002, beskin2012, pilia2016} and hence a portion of the pulsar jet may be visible off-axis, with this portion increasing with time \citep{cutler2002}. However, as shown previously, it will not be observable for viewing angles close to the orbital plane due to absorption by the ejecta from the merger process.

\subsubsection{Repeating FRBs}
\label{sec:repeatFRB}

Assuming the total mass of the magnetar remnant is less than the maximum mass allowable for a neutron star \citep[which is dependent upon the nuclear equation of state;][]{lasky2014}, the product of the compact binary merger would be a stable millisecond magnetar. As highlighted by \cite{metzger2017}, for the collapsar progenitors of long GRBs, a stable millisecond magnetar is an excellent contender for the progenitor of the repeating fast radio burst \citep[FRB 121102;][]{spitler2016}. \cite{metzger2017} illustrate that both the dipole spin down energy and the energy contained in the magnetic field of the magnetar remnant are sufficient to power repeating FRBs. However, in the case of long GRBs, the emission cannot be seen until $\sim$100 years after the explosion due to the opacity of the supernova remnant. In the compact binary merger scenario, the kilonova remnant is likely to not be isotropic (densest in the orbital plane) and hence, it may be possible to see repeating bursts along the rotation axis from the time of the magnetar formation until the rotational or magnetic energy reserves are depleted.

\subsubsection{Magnetar collapse}
\label{sec:collapse}

Some of the magnetars formed via the merger of two neutron stars are going to be too massive to be able to support themselves against collapse into a black hole. The collapse time will depend on the mass of the unstable magnetar, the initial spin and spin-down rate, and the neutron star equation of state \citep[e.g.][]{ravi2014}. \cite{falcke2014} postulate an isolated neutron star collapsing to a black hole will produce a short duration burst of coherent radio emission, comparable to an FRB, due to magnetic reconnection of the magnetosphere. \cite{zhang2014} extended this idea to predict a burst of coherent radio emission at the end of the X-ray plateau phase (phase (d) in Figure \ref{fig:CohModels}).

The amount of magnetic energy, $E_B$, released via this reconnection event is predicted to be 
\begin{eqnarray}
E_B = 1.7\times 10^{47} B_{15}^2 R_6^{3} ~{\rm erg}. \label{eqn:zhang_energy}
\end{eqnarray}
\citep{zhang2014}.
This energy is expected to be released rapidly, with an estimated timescale of $\tau=1$ ms \citep[as observed for FRBs;][]{falcke2014}. Assuming some of this energy is converted to coherent radio emission with an efficiency $\epsilon$, we can predict a bolometric luminosity, $L$, of:
\begin{eqnarray}
L = \frac{\epsilon E_B}{\tau} ~{\rm erg ~s}^{-1} \label{eqn:zhang_lumin}
\end{eqnarray}
In order to compare this theoretical value to the observations, we assume that the luminosity, as a function of the frequency, can be described as a power law:
\begin{eqnarray}
    L_{\nu} = c \nu^{\alpha},
\end{eqnarray}
where c is a constant, and the bolometric luminosity is given by:
\begin{eqnarray}
    L = \int_{\nu_p}^{\infty} L_{\nu} d\nu = \int_{\nu_p}^{\infty}  c \nu^{\alpha} d\nu = c\left[  \frac{\nu^{(\alpha +1)}}{(\alpha +1)} \right]_{\nu_p}^{\infty}.
\end{eqnarray}
As there is no emission below the plasma frequency, we can set $\nu_p$ as the lower limit in this integral. In the case where $\alpha < -1$, consistent with known coherent radio emission (e.g. pulsars), we note that as $\nu \rightarrow \infty$ then $\nu^{(\alpha + 1)} \rightarrow 0$. Therefore, we can show that the bolometric luminosity is:
\begin{eqnarray}
    L = -c  \frac{\nu_p^{(\alpha +1)}}{(\alpha +1)} = \frac{\epsilon E_B}{\tau} \\
    \rightarrow c = -\frac{\epsilon E_B}{\tau} (\alpha +1) \nu_p^{-(\alpha +1)},
\end{eqnarray}
where the observed flux density (as a function of observing frequency) is:
\begin{eqnarray}
    f_{\nu} = \frac{c (1+z) \nu_{obs}^{\alpha}}{4 \pi D^2} \nonumber \\ 
    f_{\nu} = -\frac{10^{-23} \epsilon E_B}{ 4 \pi D^2 \tau} (\alpha +1) \nu_p^{-(\alpha +1)} ~ \frac{\nu_{obs}^{\alpha}}{(1+z)} ~{\rm Jy},
\end{eqnarray}
note the factor $(1+z)$, because the observing frequency is redshifted for cosmological events. When also taking into account that the images are longer duration than the radio burst, we can show the flux density is reduced to (see Equation \ref{eqn:obsFlx}):
\begin{eqnarray}
    f_{\nu} = -\frac{10^{-23} \epsilon E_B}{ 4 \pi D^2 \tau} (\alpha +1) \nu_p^{-(\alpha +1)} ~ \frac{\nu_{obs}^{\alpha}}{(1+z)} \frac{\tau}{t_{\rm int}} ~{\rm Jy},
\end{eqnarray}

\section{Radio Telescope Capabilities}
\label{sec:telescopes}

In order to test these theoretical models predicting coherent radio emission, we require rapid radio observations of events associated with compact binary mergers. In this section, we outline the capabilities of some of the current and up-coming low frequency radio facilities that are capable of searching for this emission.

\subsection{MWA}
\label{sec:MWA}

The MWA has recently undergone its Phase II upgrade, which includes an extended baselines configuration where the maximum baseline length has been increased from 2.8 to 5.3\,km. 
While MWA Phase I reached the classical and sidelobe confusion limit in a 2\,min snapshot, the improved image resolution of the extended configuration allows for an order of magnitude improvement in the image sensitivity over longer integration times \citep{wayth2018}. 
The sensitivity of the MWA is also dependent on the pointing and frequency of the observation so for a 2\,h observation, the expected thermal noise for MWA is between 1 and 3\,mJy between 100 and 200 MHz, depending on the declination \citep{wayth2015}.

The MWA rapid-response mode has also been updated since the GRB 150424A triggered observation \citep[][see Section \ref{sec:GRB150424A}]{kaplan2015}, with the new system detailed in Hancock et al. (submitted). The upgraded system triggers on alerts transmitted via the Virtual Observatory Event standard \citep[VOEvent;][]{seaman2011} and can be on-target and recording data within $6-14$\,s following the alert. This makes MWA the ideal instrument for probing the emission models in phases (a) and (b) as shown in Figure \ref{fig:CohModels}. The active GRB program continues to trigger on both the \textit{Swfit} Burst Alert Telescope and the \textit{Fermi} Gamma-ray Burst Monitor (GBM)-detected GRBs, integrating for a total of 30\,mins. 
The VOEvent parsing strategy is optimised to prioritise GRBs that are more likely to be short. The rapid-response system can activate either the standard correlator mode (temporal and spectral resolution of 0.5\,s and 10\,kHz, respectively) or using the Voltage Capture System \citep[VCS;][]{tremblay2015}, which has a native temporal resolution of $100\,\mu\,s$. 

Sensitivities on the native MWA correlator temporal resolution of 0.5\,s can be better than 1\,Jy\,beam$^{-1}$ depending on the pointing and observing frequency (usually between $100-200$\,MHz). For example, image plane de-dispersion has already been employed for MWA FRB blind searches \citep{tingay2015} and MWA shadowing observations of ASKAP FRBs \citep{sokolowski2018}. This technique involves creating 1.28 MHz images on the shortest (0.5\,s) timescales and then stacking based on a range of expected dispersion measures, which can reach $1\sigma$ noise limits between 0.2 and 7\,Jy\,beam$^{-1}$ \citep{sokolowski2018}.  

More sensitivity can be afforded for the detection of prompt emission by triggering with the MWA VCS due to its much higher temporal resolution. For example, if we know the position of the GRB to within the MWA synthesised beam, we can conduct a coherent beam-formed single pulsed search. The corresponding sensitivity ($1\sigma$ noise) can be calculated using the radiometer equation,

\begin{eqnarray}
    \sigma_{\nu} = \frac{SEFD}{\sqrt{n_{p} t_{int} \Delta \nu}} ~{\rm Jy},
\end{eqnarray}

where $n_{p}$ is the number of sampled polarisations (2 for MWA), $t_{int}$ is the integration time, which we set to 1\,ms, $\Delta \nu$ is the bandwidth (30.72\,MHz for MWA), and SEFD is the system equivalent flux density, which we assume to be $\sim3300$\,Jy \citep[estimated using MWA VCS observations of PSR\,J1107--5907;][note this value is highly direction dependent]{meyers2018}. This results in a $3\sigma$ detection threshold of 40\,Jy at 154 MHz. 
However, incoherent beam-forming may be necessary if 
the SGRB is not localised to within a synthesised beam. If we take the median SEFD of $\sim22000$\,Jy \citep[calculated using median parameters from MWA pulsar studies at 185\,MHz derived by][]{xue2017}, then the $3\sigma$ detection threshold of an incoherent beam search is 270\,Jy \citep[which corresponds to 460\,Jy at 150\,MHz assuming the system temperature increases towards low frequencies due to extended Galactic continuum emission according to $\propto \nu^{-2.6}$;][]{lawson1987}. 
Once again, as sky dominates the system noise, the sensitivity is pointing dependent and could therefore range between $80-4000$\,Jy ($140-6800$\,Jy at 150\,MHz).

\subsection{LOFAR}
\label{sec:LOFAR}

In late 2017, the LOFAR rapid-response triggering system, known as the LOFAR Responsive Telescope, was commissioned. On receiving a transient alert, using a similar VOEvent system to the MWA and built using the 4 Pi Sky tools \citep{seaman2011,staley2016}, LOFAR Responsive Telescope triggers all the High Band Array antennas (HBA; $120-168$\,MHz) in the Dutch LOFAR stations at the central frequency of 144\,MHz with a temporal and spectral resolution of 1\,s and 3\,kHz, respectively. The LOFAR Responsive Telescope can respond within 3--5 mins, notably slower than the MWA due to current software and hardware constraints. Recently the LOFAR Responsive Telescope triggered on the long-duration gamma-ray burst (GRB) 180706A within 5\,mins of receiving the \textit{Swift}-BAT alert, obtaining a 2\,h integration (Rowlinson et al. in prep). 

Even with projected response times as fast as 3\,mins, the LOFAR Responsive Telescope is not quick enough to probe coherent emission that may be produced immediately prior or during the merger (i.e. those models from phases (a) and (b) as shown in Figure \ref{fig:CohModels}). However, the LOFAR array activated by the Responsive Telescope is far more sensitive than MWA, and can reach $3\sigma$ sensitivities potentially as low as 0.2\,mJy in 2 hours, with an instantaneous $3\sigma$ sensitivity (1\,s integration) of $\sim20$\,mJy.\footnote{These sensitivities are calculated using the LOFAR Image noise calculator; \href{https://support.astron.nl/ImageNoiseCalculator/sens2.php}{https://support.astron.nl/ImageNoiseCalculator/sens2.php}} It is therefore the most sensitive low frequency radio telescope to search for persistent dipole emission predicted to be emitted by a stable or unstable magnetar remnant during phase (c), as shown in Figure \ref{fig:CohModels}.

As with the MWA, image plane de-dispersion strategies are likely to improve the sensitivity of the LOFAR searches for short duration coherent bursts such as those during Phase (d) in \ref{fig:CohModels}. Finally, although not tested to date, LOFAR is capable of also triggering beam-formed observations alongside the deep imaging observations and this will significantly improve the sensitivity to millisecond duration events. Recently, \cite{houben2019} conducted observations of FRBs detected at higher frequencies using LOFAR. As part of their analysis, they determined that the beam-formed $1\sigma$ sensitivity of LOFAR is 4 Jy at an integration time of 1.3 ms for a single, coherent, tied-array beam with 78 MHz of bandwidth \citep{houben2019}.

\subsection{AARTFAAC}
\label{sec:AARTFAAC}

LOFAR also has an all-sky radio monitor known as the Amsterdam ASTRON Radio Transient Facility and Analysis Center (AARTFAAC), which runs as a parallel backend when the LOFAR Low Band Array (LBA; 10--90\,MHz) is operating \citep{prasad2014,prasad2016}. 
This facility samples the sky on 1\,s timescales using the inner-most 6  LOFAR core stations and 16 subbands (AARTFAAC-6; a total of 3.12\,MHz of processed bandwidth centred around 60\,MHz), with plans to upgrade to AARTFAAC-12 with the addition of 6 more core stations and another 16 subbands. The current AARTFAAC-6 configuration can reach a $3\sigma$ sensitivity of $<63$\,Jy over 90\% of the Northern Hemisphere \citep{kuiack2019}. Doubling both the number of antennas and the bandwidth for AARTFAAC-12, will lead to a typical $3\sigma$ sensitivity of 3.3\,Jy (private communication).

\subsection{LWA1}
\label{sec:LWA1}

The first station of the Long Wavelength Array (LWA1) operates in the 10--88\,MHz frequency range and has two all-sky modes, the transient-buffer narrow and the transient-buffer wide \citep{yancey2015}, the latter of which allows for continuous data recording with a bandwidth of 70\,kHz, which is tunable within the observing band. LWA1 is also equipped with the Prototype All Sky Imager (PASI) backend, which performs all-sky imaging in near real time on the native time resolution of 5\,s  \citep{obenberger2015}. In this all-sky mode, LWA1-PASI serendipitously observes GRBs, 34 of which have been presented in \citep{obenberger2014}, resulting in $3\sigma$ sensitivities of 204, 195, and 210\,Jy at 37.9, 52.0 and 74.0\,MHz, respectively. However, the limited bandwidth means that it is difficult to constrain the DM of a detected pulse, with the 5\,s integrations also reducing PASI's sensitivity to shorter-timescale prompt signals. 

However, LWA1 has had two rapid-response triggering modes commissioned, including the Burst Early Response Triggering system and the Heuristic Automation for LWA1 (HAL) system, the latter of which can respond to triggers within 2\,mins \citep{yancey2015}. These modes trigger the full LWA1 beam facility, which has a bandwidth of 19.6\,MHz and up to four pointed synthesised beams that can be used to tile portions of large positional uncertainty regions, such as those of \textit{Fermi}-detected GRBs and gravitational wave events detected by aLIGO/Virgo. While the additional dispersion delay afforded by the lower observing frequency of LWA1 means there is more time to repoint the telescope at a transient before any prompt radio emission arrives, the 2\,min response time of HAL means that it will not be able to sample prompt, coherent emission produced prior or during the merger from the SGRB population below a DM of 200\,pc\,cm$^{-3}$ \citep[between a redshift of 0.1 and 0.2;][i.e. phases (a) and (b) shown in Figure \ref{fig:CohModels}]{yancey2015}. 

\subsection{OVRO-LWA}
\label{sec:OVRO}

The Owens Valley Radio Observatory Long Wavelength Array (OVRO-LWA) is another example of an all-sky radio monitor, which observes between 27--84\,MHz with a temporal and spectral resolution of 13\,s and 24\,kHz, respectively. OVRO-LWA is equipped with a 24\,h transient buffer and on receiving an alert, can allow for visibilities to be obtained both before and after an event, which eliminates the need for low-latency transient notices.
Both \citet{anderson2018} and \citet{callister2019} have used the transient buffer to search for coherent emission from both the short GRB 170112A (see Section~\ref{sec:170112A}) and the binary black hole merger GW170104, respectively, by constructing dedispersed time-series of the dynamic spectrum over a wide range of DMs on 13\,s timescales. Within 95\% of the GW170104 localisation, \citet{callister2019} obtained a median 95\% upper-limit of 2.4\,Jy on 13\,s timescales but the sensitivity significantly degrades with decreasing elevation. The full-bandwidth limit obtained for GRB 170112A is more constraining given that \textit{Swift} localised this event to within the OVRO-LWA synthesised beam, resulting in a $3\sigma$ flux density limit of 4.5\,Jy. There are future plans to improve the voltage buffering capabilities of OVRO-LWA to have a higher temporal resolution, which will make it more sensitive to short timescale transients \citep{callister2019}.

\section{Application to observations of binary mergers}
\label{sec:application}

In this Section, we consider the tightest constraints currently obtained for coherent radio emission associated with compact binary mergers, focusing on the individual events GRB 150424A \citep{kaplan2015} and GRB 170112A \citep{anderson2018}.

\subsection{GRB 150424A}
\label{sec:GRB150424A}

The follow-up observations of GRB 150424A provide the best constraints to date for coherent radio emission as we have both the X-ray light curve and very rapid triggered radio follow-up observations with the MWA \citep{kaplan2015}. The X-ray light curve shows clear evidence of energy injection potentially generated by a newly formed magnetar that enables us to make model informed predictions of the magnetic field strength and initial spin period of this remnant.

\subsubsection{Properties}
\label{sec:propGRB}

GRB 150424A was detected by the Burst Alert Telescope \citep[BAT;][]{barthelmy2005} onboard the {\it Swift} satellite, at 07:43:10 UT on 2015 April 24 \citep{beardmore2015}. The prompt gamma-ray emission  consisted of a short hard spike of duration 0.5 seconds followed by a soft, low flux tail lasting for $\sim$100 seconds \citep{barthelmy2015, norris2015}. 

\begin{figure}
\centering
\includegraphics[width=0.45\textwidth]{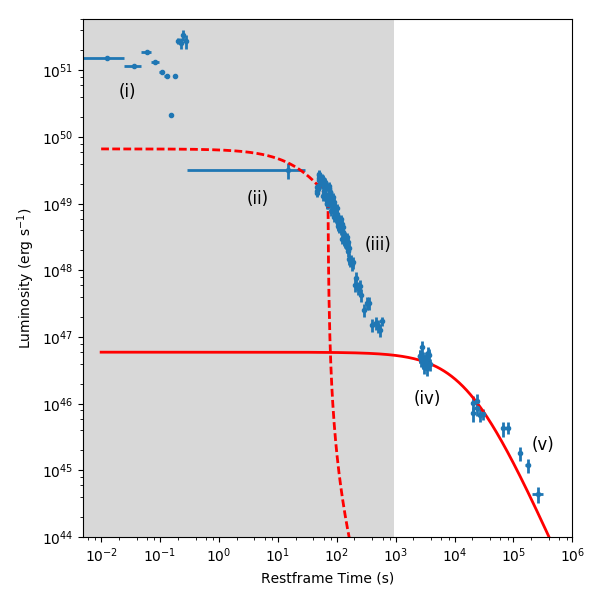}
\caption{The restframe BAT-XRT lightcurve of GRB 150424A, assuming a redshift of 0.7, where the BAT and XRT data are plotted in blue. The different phases are: (i) prompt GRB emission phase, (ii) extended emission phase, (iii) steep decay phase, (iv) plateau phase and (v) is the shallow decay phase. The dashed line shows the fit to the magnetar central engine assuming the extended emission phase is powered by the magnetar spin down phase \citep[][]{lu2015} and the solid line assumes the late-time plateau phase is powered by the magnetar spin down phase \citep[][]{rowlinson2013}. The grey shaded region shows the restframe equivalent time of the MWA observations at $\nu = 132.5$ MHz, assuming a dispersion delay of $\tau \sim 200$ s.}
\label{fig:XRTlc}
\end{figure}

After 88 seconds, the X-ray Telescope \citep[XRT;][]{burrows2005} started observations of the BAT location and detected a bright, fading X-ray counterpart \citep{melandri2015}. The X-ray emission consists of the prompt emission (i) extended emission (ii) followed by a steep decay phase (iii), a late-time plateau (iv) and shallow decay phase (v; as shown in Figure \ref{fig:XRTlc}). Optical and radio counterparts were identified in follow-up observations \citep{perley2015, fong2015}.

This GRB is 5 arcsec offset from a nearby spiral galaxy at a redshift of 0.3 \citep{castro2015}. However, there is a faint underlying extended source at the location of the GRB which may be at a redshift of $>$0.7 and is likely to be the host galaxy of the GRB \citep{tanvir2015}. Given current published observations, the redshift of this GRB remains uncertain and in this paper we assume a redshift of 0.7, consistent with the faint extended source and the average redshift of SGRBs \citep{rowlinson2013}.

MWA triggered on GRB 150424A 26 seconds following its initial detection by {\it Swift}, and conducted a 30 minute observation \citep{kaplan2015}. We focus on the most constraining limits, which were at an observing frequency of 132.5 MHz with a bandwidth of 2.56 MHz. These MWA observations, covering the entire duration of the extended X-ray emission observed by BAT and XRT (phases (i)--(iii) in Figure \ref{fig:XRTlc}), and reached a typical sensitivity of 3 Jy at 132 MHz in 4 second snapshot images. These observations were not in an optimal mode, more typical constraints are expected to be 0.1--1 Jy on 10s timescales \citep{kaplan2016}.

Although we do not know the redshift of GRB 150424A, we can obtain an approximate upper limit on the DM using Equation \ref{eqn:DM} and the adopted value of $z=0.7$. Using Equation \ref{eqn:DM} we obtain DM$<$840 pc cm$^{-3}$, and from Equation \ref{eqn:disp_delay} a maximum dispersion delay of $\sim$200\,s at 132 MHz.

Using Equation \ref{eqn:plasFreqRel}, we can calculate the co-moving plasma frequency to determine whether coherent radio emission is likely to escape the region surrounding GRB 150424A. Substituting in the peak luminosity of GRB 150424A ($\sim 10^{51}$\,erg\,s$^{-1}$) and typical values for SGRBs ($\Gamma\sim1000$ and a blast wave radius of 10$^{17}$\,cm), this corresponds to a co-moving plasma frequency of $\nu_p({\rm GRB150424A}) \simeq 0.12$ MHz. This is much lower than the MWA observing band, so we assume that coherent radio emission can escape along the jet axis of GRB 150424A.

\subsubsection{Central engine activity inferred from X-ray light curves}
\label{sec:150424Xray}

GRB 150424A has two clear X-ray signatures of ongoing energy injection into the system from the central engine; the extended emission phase and the late-time plateau. 

To apply this magnetar model to GRB 150424A, we obtained the 0.3--10 keV observed flux X-ray light curve, combining the BAT and XRT data, from the Burst Analyser page of the UK Swift Science Data Centre website \citep{evans2007, evans2009, evans2010}. In order to obtain the 1--10000 keV rest-frame light curve, we utilise a k-correction \citep{bloom2001} and assume a redshift of 0.7, the typical SGRB redshift. The rest-frame light curve is plotted in Figure \ref{fig:XRTlc} and was fitted in {\sc QDP} (Quick and Dandy Plotting from the standard {\sc HEASoft} tools, version 6.13) using equations \ref{eqn:B}, \ref{eqn:P} and \ref{eqn:Lumin}. In the following, we consider two interpretations of the magnetar central engine spin down regime: that it occurred during the X-ray extended emission versus during the X-ray plateau phase. 

First we assume that the extended emission is powered by the dipole spin-down phase (phase (ii) in Figure \ref{fig:XRTlc}) with the collapse of the magnetar into a black hole resulting in the end of the extended emission phase \citep[following the interpretation of ][]{lu2015}. We find that the steep decay following the extended emission (phase (iii) in Figure \ref{fig:XRTlc}) is inconsistent with the simple curvature effect expected to follow the collapse of the magnetar into a black hole \citep{rowlinson2010,rowlinson2013}, leading to a very poor fit to this model. We note that \cite{lu2015} suggested that the decay phase is shallower than that obtained in this modelling, so we will continue to adopt the magnetar parameters obtained from this fitting in this paper. However, due to the poor fit to the data, the collapse time is unconstrained. We trialled different collapse times in the range 70--100 seconds, and determined that the spin period is reasonably well constrained to $\sim 5.4$ ms and the magnetic field strength is in the range $\sim 3$--$5 \times10^{16}$ G at a redshift of 0.7. A typical magnetar fit is shown using the dashed lines in Figure \ref{fig:XRTlc}.  From here onwards, we will refer to this scenario as the unstable magnetar model.

Second we assume that the late time plateau phase in the X-ray light-curve is powered by the dipole emission from a stable magnetar central engine \citep{rowlinson2013} and the extended emission originates from a propellering phase as material attempts to accrete onto the magnetar \citep[the model proposed by ][]{gompertz2013,gompertz2014}. We obtain a good fit of the magnetar model to the late time plateau phase (phase (iv) in Figure \ref{fig:XRTlc}), which is shown as the solid line in Figure \ref{fig:XRTlc}. Correcting for the expected range in the beaming and efficiency ratio ($f = 3.45\pm0.29$), we find a magnetic field of $[4.30^{+0.56}_{-0.50}] \times 10^{15}$ G and a spin period of $[10.18^{+0.75}_{-0.72}]$ ms at a redshift of 0.7. From here onwards, we will refer to this scenario as the stable magnetar model.

These two models represent the two different interpretations of the X-ray extended emission and X-ray plateau within the magnetar model and hence provide different magnetar parameters. The magnetic fields and spin periods for both versions of the model are consistent with the values expected for SGRBs \citep{rowlinson2013,lu2015}. In the following sections, we will utilise both of these magnetar fits to predict the coherent low frequency radio emission expected from a magnetar central engine that may have powered GRB 150424A.

\subsubsection{Prior to merger}

The first coherent emission model we apply to GRB 150424A is that proposed by \cite{lipunov1996} and \cite{lyutikov2013}, outlined in Section \ref{sec:allignB}.

\begin{figure}
\centering
\includegraphics[width=0.45\textwidth]{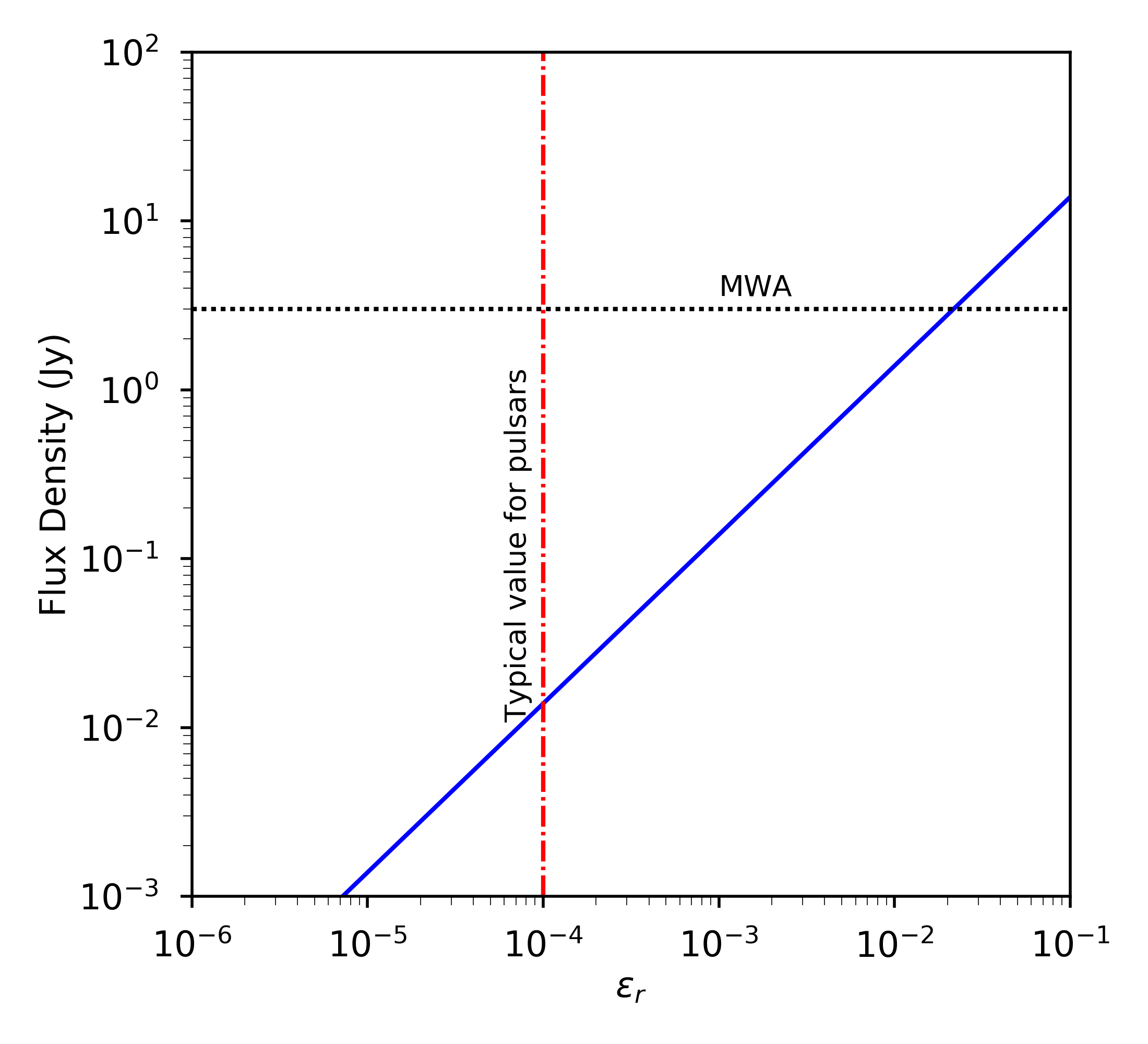}
\caption{This figure shows the predicted flux densities for GRB 150424A as a function of the efficiency for the coherent (pulsar-like) emission from the alignment of magnetic fields during a binary merger (blue solid line), using the model derived by \citet{lyutikov2013} and assuming an observing frequency of 132.5 MHz. The red dot-dashed line shows the typical efficiency observed for known pulsars \citep[e.g.][]{taylor1993b} and the black dotted line shows the flux density limit attained by the MWA on 4 second timescales for GRB 150424A \citep{kaplan2015}.}
\label{fig:lyutikov2013}
\end{figure}

In this case, we are considering emission at an observing frequency of 132.5 MHz \citep[the observing frequency that resulted in the most constraining MWA observations of GRB 150424A; ][]{kaplan2015} from one neutron star in the binary system just prior to the merger.  In this scenario, by substituting typical neutron star parameters into Equation \ref{eqn:lyutikov2013_2}, the flux density is given by
\begin{eqnarray}
F_{\nu} \sim  \frac{1500}{D^2} (1+z) \epsilon_r ~{\rm Jy} \label{eqn:lyutikov2013_3}
\end{eqnarray}
and we plot this as a function of $\epsilon_r$ in Figure \ref{fig:lyutikov2013} for the distance of 4.3 Gpc (the assumed distance to GRB 150424). For reference, we plot the typical value of $\epsilon_r = 10^{-4}$ for known pulsars \citep[e.g.][]{taylor1993b}. At a distance of 4.3 Gpc, it is clear from Figure \ref{fig:lyutikov2013} that the emission 
is undetectable in the MWA observations of GRB 150424A for most reasonable values of $\epsilon_r$. 

\subsubsection{During merger}

The second coherent emission model we apply is by \cite{usov2000}, outlined in Section \ref{sec:jetInteract}, in which the emission originates from the interaction between the relativistic jet and the surrounding material.

Using Equations \ref{eqn:UsovFlux1},  \ref{eqn:obsFlx} and \ref{eqn:UsovFlux2}, we can predict the flux density of the coherent emission. The gamma-ray fluence of GRB 150424A was measured by {\it Konus-Wind} to be $1.81 (-0.11,+0.11) \times 10^{-5}$ erg cm$^{-2}$, in the 20 keV -- 10 MeV energy band \citep{golenetskii2015}. The kinetic energy in SGRBs is relatively well constrained to be $10^{49}$ erg \citep{fong2015b}, i.e. $Q_{53} = 1\times 10^{-4}$, and the Lorentz factor has been observed to be $\Gamma_0 \sim$1000 \citep[for SGRB 090510;][]{ackermann2010}. Although expected to be low, the electron density of the surrounding medium is poorly known for SGRBs and is thought to be in the range $10^{-4}$--$1$ electrons cm$^{-3}$. In this analysis we take $n=10^{-2}$ and note that the flux density is weakly dependent on the density, $F_{\nu} \propto n^{\frac{1}{5}}$, leading to a factor of $\sim$2.5 uncertainty in the predicted flux densities. For comparison to the MWA limits, we assume an observing frequency of $\nu = 132.5$ MHz with a bandwidth of $\Delta\nu = 2.56$ MHz.

Utilising the magnetar parameters derived from the stable and unstable magnetar model fits to the X-ray light curve of GRB 150424A (see Section \ref{sec:150424Xray}), we can therefore predict the observable flux density of this coherent emission using Equations \ref{eqn:UsovFlux1} and \ref{eqn:UsovFlux2}. We neglect measurement uncertainties as these are typically less than the uncertainties on the unconstrained parameters, such as $n$. The main unknown parameters are the duration of the un-dispersed pulse, though expected to be less than the duration of the GRB, and $\epsilon_{B}$. The expected flux density is highly dependent upon both of these parameters, leading to orders of magnitude variation which make it difficult to predict a single expected flux density. We trial different values for these two parameters in the ranges $10^{-3} \leq \tau_r \leq 10$ seconds and $10^{-5} \leq \epsilon_{B} \leq 10^{-1}$. For each combination, we determine if the emission would be detectable in the MWA observations of GRB 150424A, with a limiting flux density of 3 Jy at 132.5 MHz in 4 second snapshot images for the two different magnetar fits given in Section \ref{sec:centralEng}.

\begin{figure}
\centering
\includegraphics[width=0.45\textwidth]{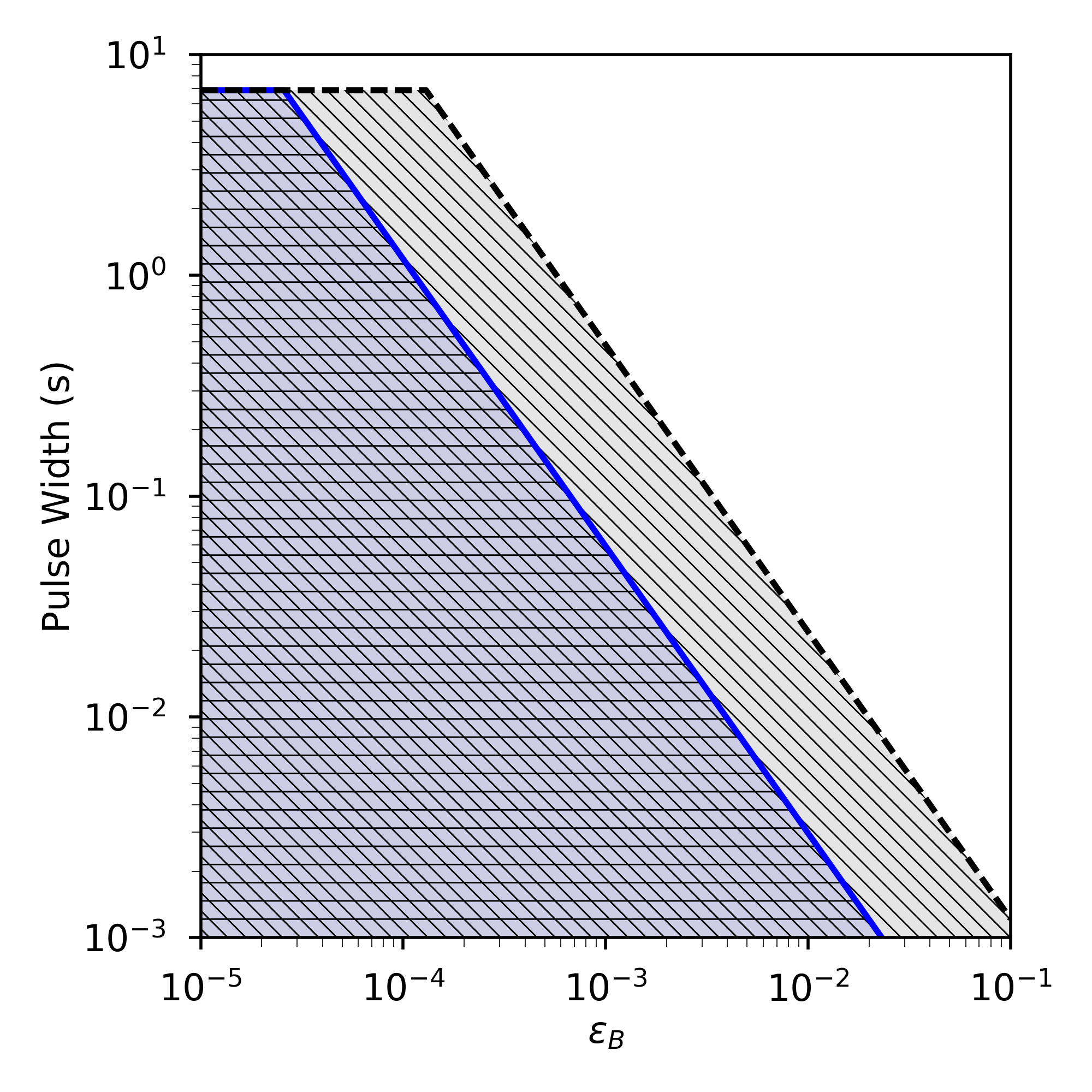}
\caption{This figure shows the region of the $\epsilon_B$ and pulse width parameter space excluded by the MWA observations of GRB 150424A. The white region is excluded, with the black dashed line representing a stable magnetar and the blue solid line representing an unstable magnetar. The flattening off of the line in the figure for low values of $\epsilon_B$ and large pulse widths marks the transition to non-dispersion limited pulses.}
\label{fig:usov}
\end{figure}

In Figure \ref{fig:usov} we plot the region excluded by the observations of GRB 150424A (white region;  i.e. where the predicted coherent emission flux would be above the MWA detection threshold for GRB 150424A and should have been detectable in the observations) and the parameter space still allowed by the observations (hatching; i.e. where the predicted coherent emission flux would be below the MWA detection threshold for GRB 150424A and would not have been detectable in the observations). Pulses with durations $\gtrsim$6 s are excluded for a redshift of 0.7. For low values of $\epsilon_{B}$ these curves flatten as the pulses are not dispersion limited. In the dispersion limited case, high values of $\epsilon_{B}$ and longer pulse durations are also excluded.

\subsubsection{Persistent emission following merger}
The third model to be considered assumes that the resulting magnetar remnant behaves like a pulsar, producing dipole radiation that is detected as persistent coherent radio emission for its lifetime, as described in Section \ref{sec:pulsar} and \cite{totani2013}.

In the case of GRB 150424A, where we propose the observed X-ray energy injection is caused by a magnetar, we can constrain this timescale to be less than $\sim 100$ s (if the magnetar collapses at the end of the extended emission phase, phase (ii) in Figure \ref{fig:XRTlc}) or $>10^{5}$ s (if the late time plateau is powered by the magnetar, phase (iv) in Figure \ref{fig:XRTlc}). 

\begin{figure}
\centering
\includegraphics[width=0.45\textwidth]{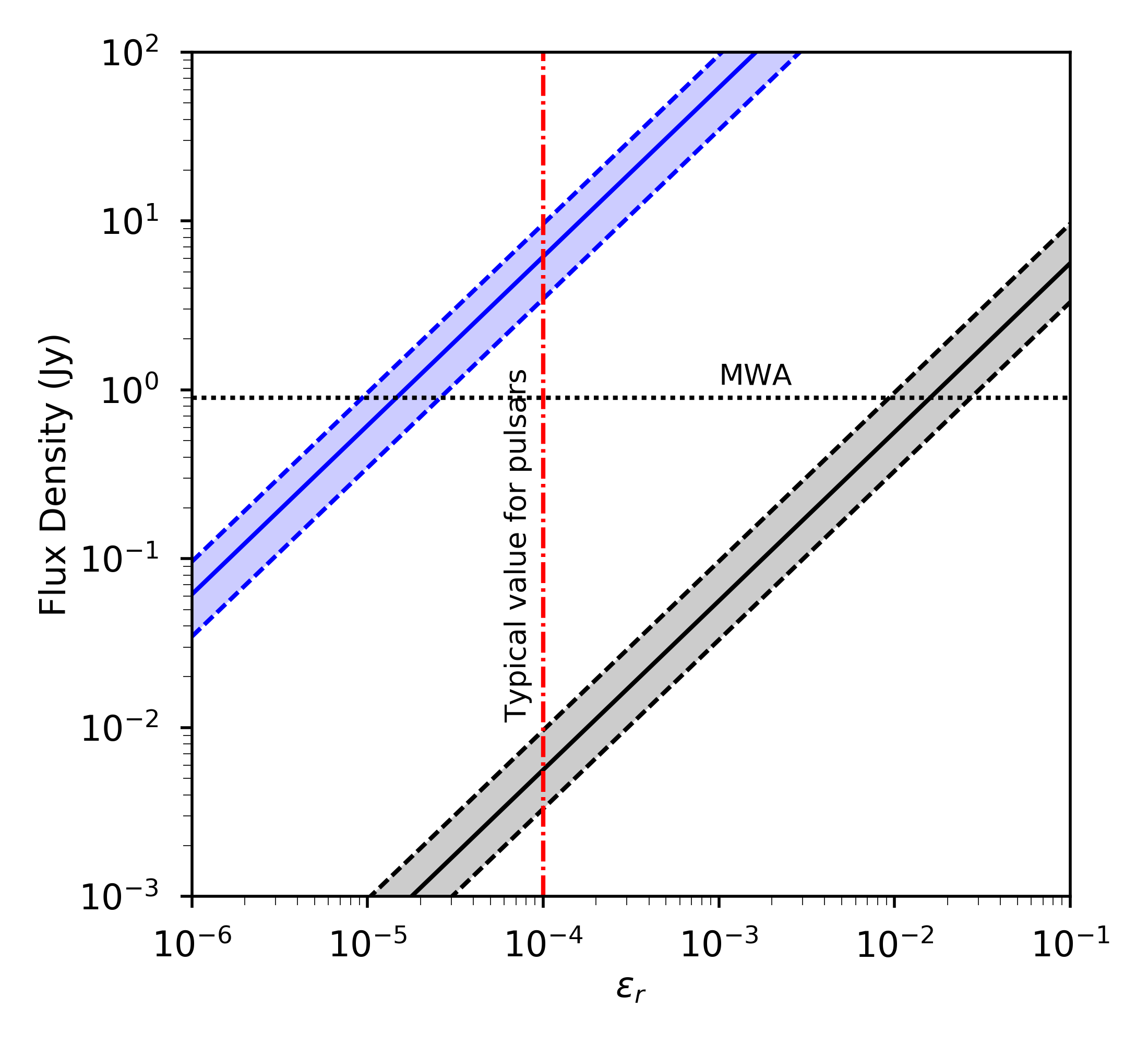}
\caption{This figure shows the predicted flux densities as a function of the efficiency for the pulsar-like emission emitted by the millisecond magnetar predicted to form via the merger of two neutron stars in Section \ref{sec:pulsar}. The solid blue line and associated shaded region show the predictions (with 1$\sigma$ uncertainty) for the unstable magnetar model outlined in Section \ref{sec:centralEng}. The black solid line and shaded region show the predictions for the stable magnetar scenario. These predictions are calculated using the model derived by \citet{totani2013} and assume an observing frequency of 132.5 MHz. The red dot-dashed line shows the typical efficiency observed for known pulsars and the black dotted line shows the $3\sigma$ flux density limit attained in the deep 30 minute observation of GRB 150424A by the MWA \citep{kaplan2015}.}
\label{fig:totani}
\end{figure}

At an adopted redshift of 0.7 corresponding to $D=4.3$ Gpc, an observing frequency of 132.5 MHz, and the magnetic fields and spin periods for the two magnetar fits outlined in Section \ref{sec:150424Xray}, we show that the expected flux densities are:
\begin{eqnarray}
F_{\nu} \simeq [6 \pm 3]\times10^{4} \epsilon_{r} ~{\rm Jy} ~{\rm (Unstable)} \label{eqn:totaniUnstable}\\ 
F_{\nu} \simeq 63^{+40}_{-23} \epsilon_{r} ~{\rm Jy} ~{\rm (Stable)} \label{eqn:totaniStable}
\end{eqnarray}
The efficiency is the remaining unknown quantity in this analysis, with pulsars typically having a value of $10^{-4}$ \citep[e.g.][]{taylor1993b}. In this analysis we take a range of values of $\eta_r$, $10^{-6}$--$0.1$, but note that the efficiency is most likely to be comparable to that of pulsars given that the emission mechanism is thought to be the same. In Figure \ref{fig:totani}, we plot the expected flux densities for the emission from GRB 150424A for the two different magnetar model fits. We note that for typical pulsar efficiencies, the MWA observations rule out that the extended X-ray emission (phase (ii) in Figure \ref{fig:XRTlc}) was powered by the spin-down from an unstable magnetar. Given that the remnant magnetar was likely stable based on our modelling in Figure \ref{fig:XRTlc}, any associated radio emission would have lasted the duration of the MWA observation and is persistent. This means that we could potentially detect this emission with long integrations rather than relying on the sensitivity of short-timescale snapshots.

At 18.02 hours after the GRB, the VLA \citep[Very Large Array;][]{thompson1980} detected a faint radio counterpart with a flux density of $\sim$31 mJy at an observing frequency of 9.8 GHz \citep{fong2015}. We can therefore calculate the contribution of persistent dipole radiation from a magnetar remnant to this VLA detection (note that we have already shown that the plasma frequency of the blast wave in the jet of GRB 150424A is sufficiently low to enable this emission to be observable). Assuming the stable magnetar fit to GRB 150424A, the magnetar will be spinning down rapidly via dipole radiation during those 18 hours after formation \citep[neglecting additional losses via gravitational wave radiation; e.g.][]{corsi2009}. Using equations 5 and 6 from \cite{rowlinson2017}, we can predict the spin period of the stable magnetar to be $8.19^{+0.61}_{-0.50}$ seconds at 18.02 hours after formation. Using equation \ref{eqn:totaniFlx}, we can predict the flux density contribution from a stable magnetar at 9.8 GHz, at 18.02 hours after formation, to be $2\times10^{-12} \epsilon_{r}$ Jy. Therefore, we do not expect to be able to detect the persistent emission at this time and the VLA detection is likely from the standard radio afterglow of the GRB.

\subsubsection{Collapse to a black hole}

Finally, we consider the scenario where the magnetar powering the extended emission in GRB 150424A produces prompt, coherent emission as it collapses to form a black hole at $\sim$100 seconds \citep[as outlined in Section \ref{sec:collapse}, Section \ref{sec:150424Xray} and ][]{zhang2014}. Assuming a snapshot integration time of 4 seconds, a pulse width of 1 ms, a redshift of 0.7 (corresponding to a distance of 4.3 Gpc), an observing frequency of 132.5 MHz, and the unstable magnetar properties calculated in Section \ref{sec:150424Xray}, we can predict the flux density of this emission. The remaining unknown parameters are the efficiency, taken to be $10^{-10} \le \epsilon \le 10^{-2}$, and the spectral index of the emission, where we take 3 representative values ($\alpha =$ -2, -3, and -4). 

\begin{figure}
\centering
\includegraphics[width=0.45\textwidth]{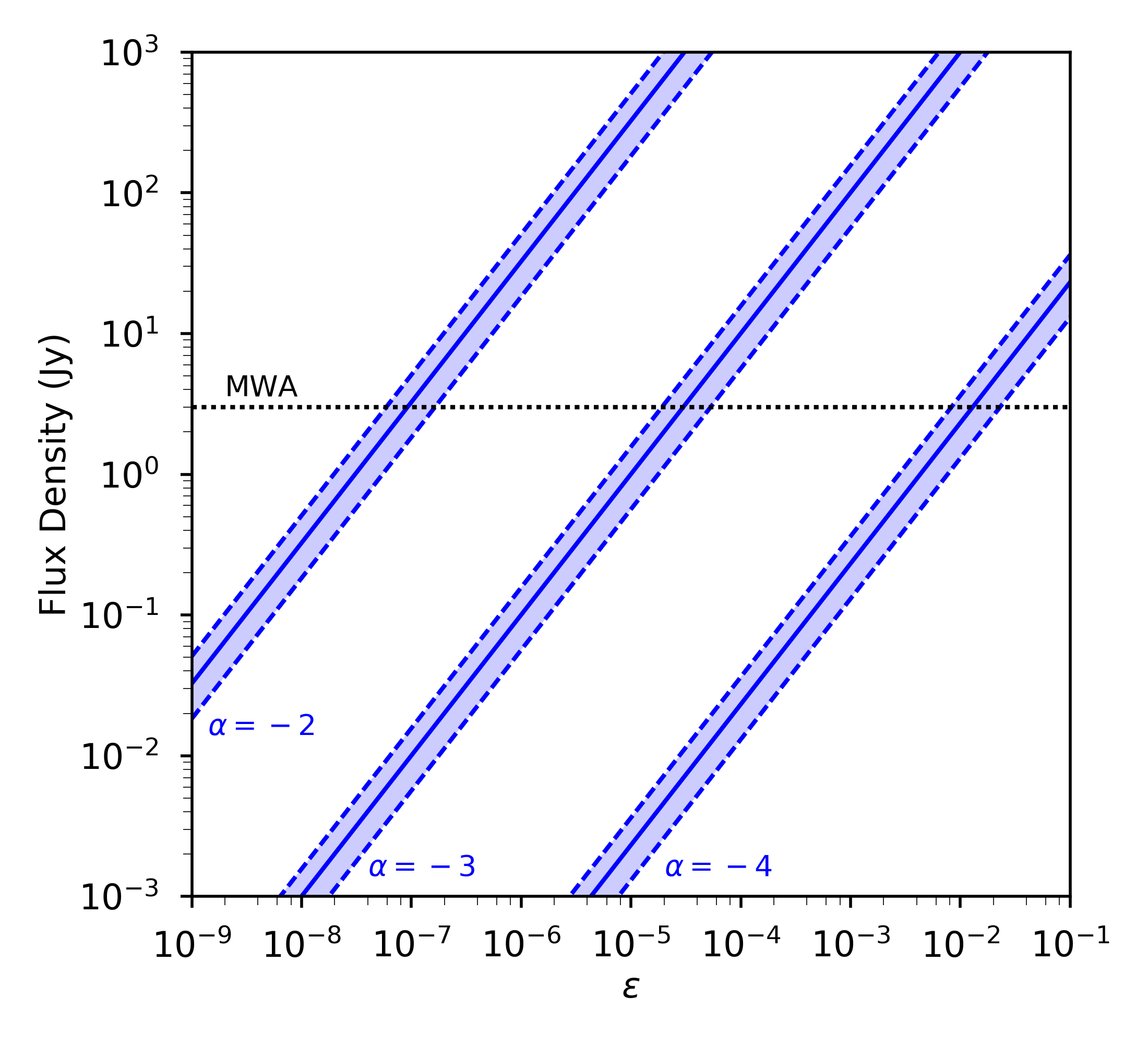}
\caption{The expected flux densities of the emission associated with a magnetar collapsing to form a black hole \citep{falcke2014,zhang2014}, assuming a pulse duration of 1 ms and a snapshot integration time of 4 s for a range of efficiencies ($\epsilon$) and spectral indicies (represented by the dashed blue lines). This emission is predicted using parameters derived from fitting the unstable magnetar model fitted to the extended emission of GRB 150424A (see Figure \ref{fig:XRTlc}), assuming a redshift of 0.7. The black dotted lines represent the sensitivity of the MWA 4 second snapshots of GRB 150424A.}
\label{fig:zhang}
\end{figure}

In Figure \ref{fig:zhang}, we plot the predicted flux densities of the coherent radio emission associated with the collapse of an unstable magnetar into a black hole for GRB 150424A as a function of efficiency, and show this for three different spectral indices. As shown by \cite{zhang2014}, and in the specific case of GRB 150424A, this emission is expected to be detectable for on-axis jet viewing angles. However, it will be more difficult to detect for off-axis jet viewing angles as the detectability depends upon the distribution of the binary merger ejecta.

The non-detection of this coherent emission in the case of GRB 150424A, despite it seeming to be detectable from Figure \ref{fig:zhang}, can be interpreted in the following ways:
\begin{enumerate}
  \item The magnetar did not collapse at the end of the extended emission phase as postulated by \cite{zhang2014,lu2015}, which would be consistent with the modelling of \cite{rowlinson2013,gompertz2013,gompertz2014} that shows the extended emission was from a propellering phase and that the late-time plateau phase was powered by the spin-down of a stable magnetar.
  \item The spectrum of the emission was very steep.
  \item This energy was not converted to coherent radio emission or the efficiency was very low.
  \item The plasma frequency of the relativistic blast wave was orders of magnitude higher than predicted, preventing the emission from escaping the system.
\end{enumerate}

This analysis shows that for future SGRBs with a known redshift, and for which the plateau phase can be clearly attributed to an unstable magnetar \citep[e.g. GRB 090515;][]{rowlinson2010}, we will be able to place very tight constraints on this emission mechanism. If the newly formed magnetar is unstable, the collapse is likely to occur within 2 hours of the merger \citep{ravi2014}.

\subsubsection{Constraints with redshift}
\label{sec:redshift}

As we do not know the redshift of GRB 150424A, we need to consider how the predictions will change if the GRB was closer or further away. In \cite{rowlinson2010}, it was clear that the fitted magnetar parameters for GRB 090515 changed as the event was modeled at higher or lower redshifts. This is because the magnetar parameters are dependent upon the luminosity and restframe duration of the plateau. Here we derive an analytic solution that describes how these parameters evolve with redshift. We first converted the redshifts into luminosity distances ($D$) using the cosmology parameters defined in the introduction\footnote{using the Ned Wright's Cosmology Calculator, \\ http://www.astro.ucla.edu/$\sim$wright/CosmoCalc.html \citep{wright2006}}. The magnetic field strength and pulsar spin period defined in equations \ref{eqn:B} and \ref{eqn:P} therefore scale as:
\begin{eqnarray}
B_{15} \propto D^{-1} (1+z) \label{eqn:Bprop} \\
P_{-3} \propto D^{-1} (1+z)^{\frac{1}{2}} \label{eqn:Pprop}
\end{eqnarray}
We tested these relationships using the fitted results obtained for GRB 090515. These relationships allowed us to scale the results at z=0.7 to a redshift range of 0.1 -- 1. In Figure \ref{fig:150424A_z_plot}, we show the predicted emission for GRB 150424A as a function of redshift for the following scenarios (as previously described in Section \ref{sec:coherent}):

\begin{figure}
\centering
\includegraphics[width=0.33\textwidth]{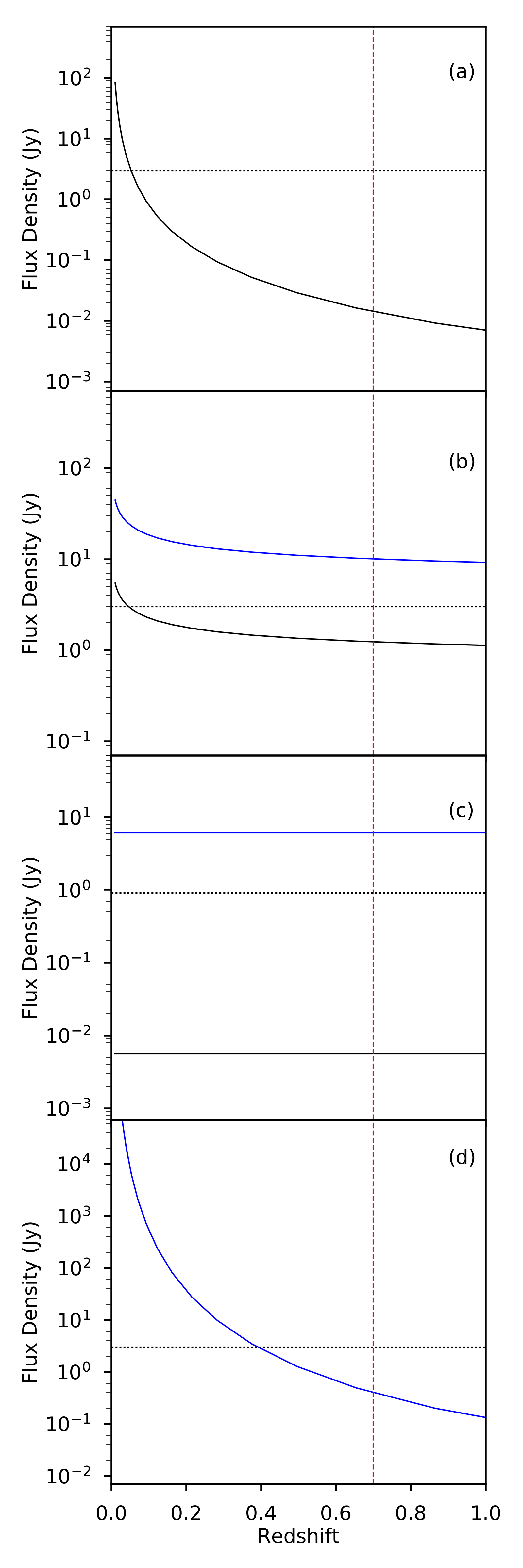}
\caption{GRB150424A currently does not have a good redshift constraint so we analytically calculate the effect on the model predictions for the flux density if the GRB is at a different redshift to the 0.7 assumed previously. a) shows the emission prior to merger and is agnostic to the merger remnant, b) shows the emission predicted during the merger  from the jet interaction with the ISM, c) shows the persistent emission associated with a newly formed magnetar and d) shows the emission associated with the collapse of the magnetar to form a black hole. The blue and black lines represent the unstable and stable magnetars, predicted from the modelling shown in Figure \ref{fig:XRTlc}, respectively. The black dotted line is the flux density limit attained by the MWA \citep[3 Jy, 4 seconds integration, for short pulse models (a, b, and d) and 0.9 Jy, 30 minute integration, for the persistent emission model (c);][]{kaplan2015} and the red dash-dotted line shows the average redshift for SGRBs, z = 0.7.}
\label{fig:150424A_z_plot}
\end{figure}

\begin{enumerate}[(a)]
   \item {\bf Prior to merger:} We predict the flux densities caused by the alignment of the magnetic fields between two neutron stars prior to merger using the model outlined in Section \ref{sec:allignB} \citep[e.g][]{lyutikov2013}. Here, in addition to the assumptions outlined previously, we assume a typical pulsar efficiency of $\epsilon_r = 10^{-4}$, and  scale the flux densities as a function of redshift by using $\frac{f1}{f2} = \left(\frac{D2}{D1}\right)^{2}$, where D1 is luminosity distance for the redshift of 0.7 and D2 represents the luminosity distances to the redshifts considered. Based on this scaling, only mergers within a redshift of 0.1 would have been detected with the MWA 4 second snapshot observations of GRB 150424A, as shown in Figure \ref{fig:150424A_z_plot}(a).
   \item {\bf Relativistic jet and ISM interaction:} The shock front caused by a magnetised relativistic jet colliding with the ISM is predicted to produce coherent radio emission (see Section \ref{sec:jetInteract}). The magnetic field at the shock front in the model by \cite{usov2000} is dependent upon the magnetic field and spin period of the newly formed magnetar. As outlined above, the spin period and the magnetic field will change for this event with redshift and hence, produce a change in brightness of the predicted flux density. Here we assume that the emission has the same duration as the prompt emission of GRB 150424A, i.e. 0.3 seconds, and we assume that $\epsilon_B = 10^{-3}$. Otherwise, we assume the same parameters as outlined in Section \ref{sec:jetInteract}. There are two candidate magnetar models, the unstable magnetar that collapses at the end of the extended emission phase (blue line) and the stable magnetar candidate (black line) as shown in Figure \ref{fig:XRTlc}. In Figure \ref{fig:150424A_z_plot}(b) it can be seen that the emission associated with the unstable magnetar would have been observable during the prompt GRB emission phase (phase (i) in Figure \ref{fig:XRTlc}) in the MWA 4 second snapshots of GRB 150424A, whereas the stable magnetar prompt emission would only be observable for redshifts $<$0.1.
   \item {\bf Persistent pulsar-like emission:} As outlined in Section \ref{sec:pulsar}, we expect the magnetar to emit in a similar way to pulsars \citep[e.g.][]{totani2013}. This emission is again dependent upon the magnetic field and spin period of the magnetar (calculated using Equations \ref{eqn:Bprop} and \ref{eqn:Pprop}). Here we assume that the efficiency is identical to that of known pulsars \citep[i.e. $10^{-4}$;][]{taylor1993b}. Interestingly, once the distance factors are properly accounted for, this model produces a constant flux density with redshift. Again, the pulsar emission associated with the unstable magnetar model would have been observable in the MWA 30 minute integration of GRB 150424A for all redshifts, however, the emission associated with the stable candidate would have been undetected as demonstrated in Figure \ref{fig:150424A_z_plot}(c).
   \item {\bf Magnetar collapse:} This model is only applicable to the unstable magnetar scenario and predicts the emission caused by magnetic reconnection as the magnetar collapses to form a black hole \citep{zhang2014}, see Section \ref{sec:collapse}. This emission is naturally dependent upon the magnetic field of the magnetar, which changes with redshift as outlined above. Here we assume a spectral index of -3 and an efficiency of $10^{-6}$, and otherwise adopt the parameters given in Section \ref{sec:collapse}. Under these assumptions, we would have only observed this emission in the MWA 4 second snapshots of GRB 150424A if it were at redshifts $<$0.4 (see Figure \ref{fig:150424A_z_plot}(d)). However, this result is highly dependent on the spectral index of the coherent emission.
\end{enumerate}

This analysis demonstrates that even with the significant uncertainties in each of the models, we can state with reasonable confidence that if GRB 150424A formed an unstable magnetar that collapsed at $\sim$100 seconds, the MWA observations would have detected the coherent radio emission associated with the prompt GRB emission phase (phase (i) in Figure \ref{fig:XRTlc}) and the newly formed magnetar (phase (ii) in Figure \ref{fig:XRTlc}). We therefore conclude that either the models and/or the assumptions are incorrect, or the unstable magnetar model is incorrect for GRB 150424A.

For the stable magnetar scenario, favoured by \cite{gompertz2013,gompertz2014}, the only detectable emission would have occurred during or prior to the prompt phase and would only be observable at low redshifts.

\subsection{GRB 170112A}
\label{sec:170112A}

The OVRO-LWA was on-sky and observing the position of GRB 170112A when it was detected by {\it Swift} BAT, obtaining simultaneous observations \citep{anderson2017}. Their searches at 27 -- 84 MHz resulted in a non-detection with a flux density limit of 650 mJy. This non-detection, which is to a deeper limits than that obtained for 150424A, is an interesting case study for this work.

GRB 170112A was an unambiguously short GRB detected by the {\it Swift} Satellite with a duration of 0.06 s \citep{lien2017}. Following a very prompt slew of the satellite, taking just 62 seconds, no X-ray emission was detected \citep{dai2017}. This non-detection is extremely unusual for Short GRBs, with only 3 other short GRBs having no detected X-ray afterglow following a prompt slew\footnote{These events could be off-axis events similar to GW/GRB 170817 where the X-ray counterpart was not detected until days after the merger.}. 

The non-detection of X-ray emission leads to the expectation that the event is either a neutron star - black hole merger or that the remnant collapsed to form a black hole within the 62 seconds it took for the satellite to slew.

If the progenitor of GRB 170112A was the merger of a neutron star and a black hole, the merger remnant would obviously be a black hole. The prompt emission mechanism outlined in Section \ref{sec:jetInteract} proposed by \cite{usov2000}, requires a highly magnetised wind from a fast rotating neutron star so does not predict coherent emission for a black hole remnant. The presence of a black hole also rules out the model proposed by \cite{lipunov1996} presented in Section \ref{sec:allignB}, as it requires the presence of two neutron stars. Therefore, the only coherent radio emission mechanisms that are expected for a neutron star and black hole system are via the excitation of plasma by gravitational waves \citep[][]{moortgat2003} or from an orbital motion battery mechanism \citep[][]{mingarelli2015}, see Section \ref{sec:othermodels}. These emission mechanisms would produce a short burst of radio emission in the final seconds of the merger, arriving at the observer up to a few minutes, for a reasonable redshift of 0.7 at the LWA observing frequencies, after the gamma-rays are detected (depending on the dispersion delay at the observing frequency).

Alternatively, GRB 170112A may be from the merger of two neutron stars where the merger remnant collapses to form a black hole within 62 seconds. If the collapse took longer than this, we would expect to observe the signature of energy injection in the X-ray lightcurve. In this scenario, all of the emission mechanisms outlined in Section \ref{sec:coherent} are possible depending upon the exact collapse time of the merger remnant, with all coherent radio emission ending when the black hole forms (a few minutes post-merger for the observer due to the dispersion delay at LWA observing frequencies). However, without a redshift measurement or the detection of X-ray emission, both of which are required to identify the nature and properties of the merger remnant (as conducted for GRB 150424A, see Section \ref{sec:150424Xray}), there are too many free parameters to tightly constrain the various emission models.

\section{Future observations of binary mergers}
\label{sec:future}

In this Section, we will consider the potential of detecting coherent radio emission from compact binary mergers in the future (from both cosmological SGRBs and gravitational wave events) and the facilities that are optimally placed to make such detections.

\subsection{Cosmological Short GRBs}
\label{sec:cosmoSGRBs}

In order to explore the detectability of other SGRBs with current radio facilities, we first need to consider the typical merger remnants formed during these events by modelling the X-ray light curves of multiple events.  As there are a wide range of parameters used to constrain the coherent radio emission expected from SGRBs, we consider if the emission could be detectable for the general population. Rapid response radio follow-up of a larger sample of these events is therefore necessary to increase the probability of capturing a system that formed a magnetar via the merger of two neutron stars. 

For many of the events detected by the {\it Swift} satellite there will have be an X-ray light curve and often a redshift, which will allow for modelling to identify energy injection signatures in the lightcurve. In the ideal scenario, the detected SGRB will have an the X-ray lightcurve and known redshift, thus reducing the number of uncertain parameters in the magnetar modelling and enabling us to tightly constrain the emission models presented in this paper. However, with the required rapid response timescales, it is not known at the trigger time if these will be available. This means that multiple triggers are required in order to increase the likelihood of detecting such an event.

\begin{figure}
\centering
\includegraphics[width=0.45\textwidth]{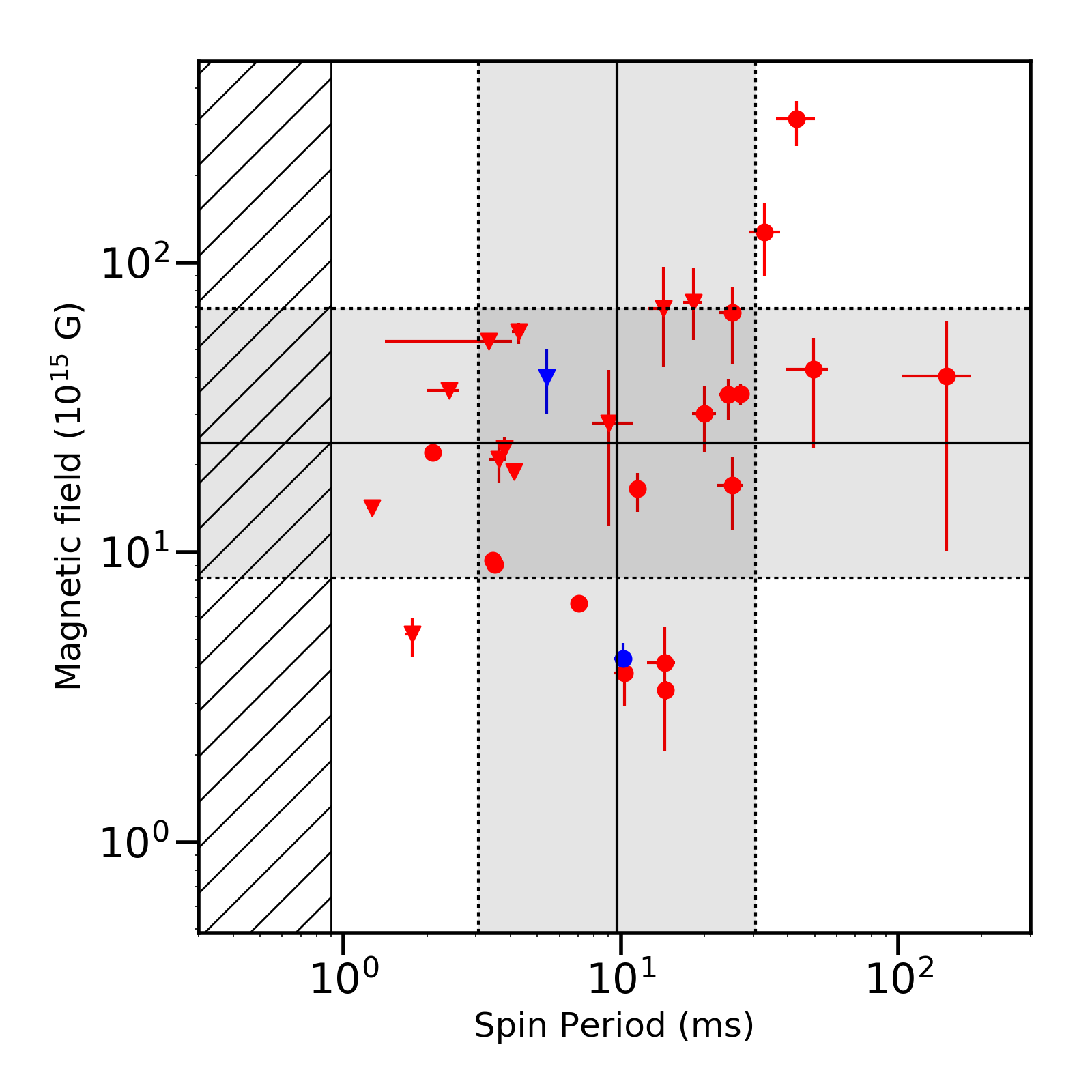}
\caption{This figure shows the magnetic field and spin periods of the SGRBs studied by \citet{rowlinson2013}, where their parameters have been adjusted to correct for beaming and efficiency using the results from \citet{rowlinson2014}. The hashed region represents the forbidden region, within which the neutron star will be broken up due to the speed of its rotation. The red circles are the SGRBs that are believed to form a stable magnetar and the red triangles are those that are thought to have an unstable magnetar. The blue circles and triangles are the results for the stable and unstable (respectively) magnetars fitted for GRB 150424. The solid black line and shaded regions represent the mean and $1\sigma$ scatter in the distributions for the magnetic field and the spin period (fitted in logarithmic space).}
\label{fig:BPplot}
\end{figure}

\cite{rowlinson2014} explored the X-ray lightcurves of the population of SGRBs detected by the {\it Swift} Satellite and found that a large number were consistent with having energy injection from a newly formed magnetar central engine. In Figure \ref{fig:BPplot}, we plot the magnetic field and spin period of the sample of magnetars fitted to their X-ray plateaus by \cite{rowlinson2013} after correcting them for the expected beaming and efficiency factor determined using the results from \cite{rowlinson2014}. GRB 150424A is included in this plot for reference using blue data points. Using the $\log_{10}$ values of the magnetic fields and spin periods, we calculate their means and the standard deviations. We define a ``typical'' magnetar, formed via the merger of two neutron stars, as having a magnetic field strength of $2.4^{+4.6}_{-1.6} \times 10^{16}$ G and a spin period of $9.7^{+20.8}_{-6.6}$ ms.

\begin{figure}
\centering
\includegraphics[width=0.30\textwidth]{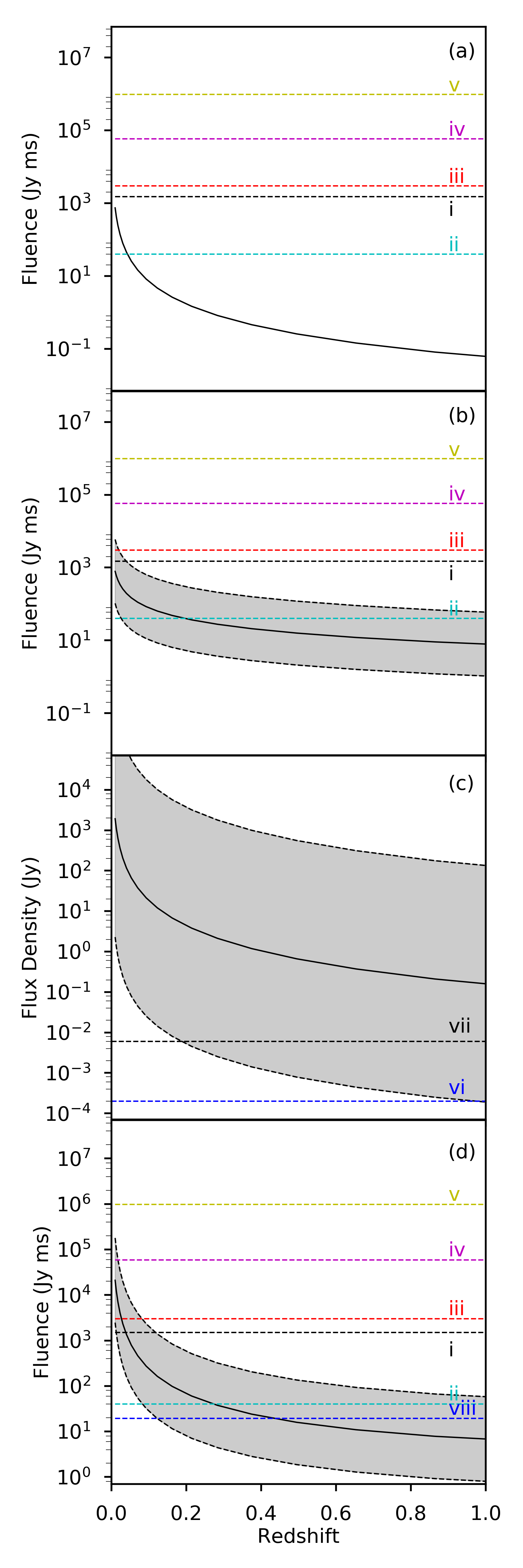}
\caption{Similar to Figure \ref{fig:150424A_z_plot}, here we show the predicted flux density or fluence (for short duration emission models, assuming a 10 ms pulse), assuming an observing frequency of 150 MHz with a bandwidth of 1 MHz, for a `typical' SGRB at different redshifts and using the mean parameters for the magnetic field and spin period shown in Figure \ref{fig:BPplot} with their associated $1\sigma$ scatters. Therefore, in Figure \ref{fig:150424A_z_plot} the magnetic field and spin period were changing with redshift to fit the observations, while here we instead assume constant values and move the magnetar to different redshifts. a) shows the emission prior to merger and is agnostic to the merger remnant, b) shows the emission predicted during the merger from the jet interaction with the ISM, c) shows the persistent emission associated with a newly formed magnetar and d) shows the emission associated with the collapse of the magnetar to form a black hole. The labeled dashed lines are the flux density limits attained by the radio telescopes considered in Section \ref{sec:telescopes}, with further details in Section \ref{sec:cosmoSGRBs} (i= MWA, ii= MWA VCS coherent, iii= AARTFAAC-12, iv= OVRO-LWA, v= LWA-PASI, vi= LOFAR deep, vii= MWA deep, viii= LOFAR 1s snapshot)} \label{fig:model_z_plot}
\end{figure}

Using the ``typical'' magnetar parameters, we predict the coherent emission expected at 150 MHz for a range of redshifts, using the coherent emission model assumptions outlined in Section \ref{sec:redshift}, which are illustrated in Figure \ref{fig:model_z_plot}. We assume a frequency bandwidth of 1 MHz for this figure,  but note that using smaller bandwidths and finer time sampling could significantly improve the sensitivity of the radio facilities to the short duration coherent emission. The short coherent pulses are assumed to have a width of 10 ms and experience zero scattering. The predicted radio emission depicted in Figure \ref{fig:model_z_plot} is also compared to the typical sensitivities of current low frequency radio telescopes that are targeting these prompt signals (see Section \ref{sec:telescopes} for details) and we discuss their capabilities in the following sections.

\subsubsection{MWA}

The MWA capabilities are outlined in Section \ref{sec:MWA} and an approximate 0.5\,s snapshot $3\sigma$ flux density and fluence limits have been included in Figure~\ref{fig:model_z_plot}. It is clear that such observations may not be sensitive enough to probe the coherent emission predicted to be associated with the initial merger (see Figure~\ref{fig:model_z_plot}\,a and \ref{fig:model_z_plot}\,b) but could potentially detect that which may be associated with the collapse of the magnetar (Figure~\ref{fig:model_z_plot}\,d), provided this occurs during the MWA 30\,minute triggered observation. Deep MWA observations (e.g. $\sim$2 hour integrations) with the Phase II extended configuration may also be sensitive enough to detect persistent coherent emission from a magnetar remnant (Figure \ref{fig:model_z_plot}\,c)

We include the MWA VCS coherent beam-formed sensitivity in Figure \ref{fig:model_z_plot}\,a, \ref{fig:model_z_plot}\,b and \ref{fig:model_z_plot}\,d, assuming a well localised event. Figure \ref{fig:model_z_plot}\,a and \ref{fig:model_z_plot}\,b shows that the MWA VCS coherent mode is the most sensitive facility for probing coherent emission produced prior and during the merger
and therefore has the highest chance of detecting such signals at redshifts up to $\sim$0.1 and $\sim$ 0.3 (for regimes a and b respectively). 
The MWA VCS is also extremely competitive for detecting prompt emission associated with the collapse of an unstable magentar into a black hole, and is sensitive to this signature over the entire explored redshift range.

\subsubsection{LOFAR}

As outlined in Section \ref{sec:LOFAR}, LOFAR is insufficiently fast to probe the earliest models but, once on source, is the most sensitive facility. This means that LOFAR Responsive Telescope triggered observations are the most competitive for constraining the later-time coherent emission mechanisms predicted for SGRBs shown in Figure~\ref{fig:model_z_plot}\,c and \ref{fig:model_z_plot}\,d. 

LOFAR is also capable of deep beam-formed observations (as explained in Section \ref{sec:LOFAR}). Assuming the described observational set-up was replicated, the $3\sigma$ fluence limit for LOFAR beam-formed observations is comparable to that of the LOFAR 1 second snapshot images shown in Figure \ref{fig:model_z_plot}\,d. In summary, LOFAR is well placed to detect the coherent radio emission from these systems on timescales $\gtrsim 3$ mins.

\subsubsection{AARTFAAC}

While AARTFAAC (as described in \ref{sec:AARTFAAC}) has the potential to be on-sky at the moment of an SGRB, it is unlikely that AARTFAAC-12 will be sensitive enough on short timescales to probe for associated prompt emission for the majority of events. We note that for the nearest events (such as those associated with gravitational wave detections), and given the model uncertainties, AARTFAAC-12 still holds the potential for detecting the prompt coherent emission associated with the collapse of an unstable magnetar (see Figure~\ref{fig:model_z_plot}b).

\subsubsection{LWA1}

Similar to AARTFAAC, LWA1-PASI (described in Section \ref{sec:LWA1}), is capable of being on-sky at the time of a SGRB. However, as shown in Figures \ref{fig:model_z_plot}\,a,  \ref{fig:model_z_plot}\,b and \ref{fig:model_z_plot}\,d, LWA1-PASI is insufficiently sensitive to probe for prompt, coherent emission associated with compact binary mergers.

As stated in Section \ref{sec:LWA1}, LWA1 has two rapid response triggering modes providing higher time resolution data. The impressive temporal resolution of LWA1 \citep[microsecond timescales; e.g.][]{stovall2015}, will make these triggered observations much more sensitive to the predicted short-timescale (millisecond) prompt radio signals shown in Figure~\ref{fig:model_z_plot}. 

\subsubsection{OVRO-LWA}

While the sensitivity of OVRO-LWA (see Section \ref{sec:OVRO}) provides constraints for the early emission models, the long integration time of 13\,s significantly reduces the OVRO-LWA sensitivity to prompt radio signals as shown in Figures \ref{fig:model_z_plot}\,a, \ref{fig:model_z_plot}\,b and \ref{fig:model_z_plot}\,d.

\subsubsection{Overview of facilities and detectability of emission}

The predicted prompt radio emission from SGRBs illustrated in Figure  \ref{fig:model_z_plot}\,a and \ref{fig:model_z_plot}\,b demonstrate that only the MWA VCS mode is capable of detecting these signals, which is due to its fast response and millisecond time-scale resolution, but only for systems $z\lesssim$0.2. 
Assuming a flat spectrum, the whole sky monitoring systems (OVRO-LWA and LWA-PASI) are either insufficiently sensitive or their snapshot integration times are too long to detect any coherent radio emission associated with SGRBs. AARTFAAC-12 is the only exception to this, having a small chance of detecting prompt, coherent radio emission from very nearby events that are detected by aLIGO/Virgo via their gravitational wave emission.
 
 For the prolonged pulsar-like emission shown in Figure~\ref{fig:model_z_plot}\,c, while it is clear that even without relaxing the model assumptions there is considerable variation in the predicted emission, there is a strong possibility of it being detectable using deep integrations with MWA and LOFAR. 
 The emission expected when the magnetar collapses to form 
 a black hole (see Figure~\ref{fig:model_z_plot}\,d) is likely to be bright, with most of the facilities being able to detect this at low redshifts. The 1 second snapshot LOFAR images, LOFAR beam-formed observations and MWA VCS observations will be key for detecting this emission. This result remains strongly dependent upon the spectral index of the emission so a large number of triggered low-frequency radio observations will be required to constrain this emission.

\subsection{Compact binary mergers detected via gravitational waves}
\label{sec:GWevents}

Over the coming years, we expect to detect more compact binary mergers using gravitational wave observatories. These events are going to be significantly closer than the population of cosmological SGRBs and will therefore produce significantly higher flux coherent radio emission (see redshifts $<$0.1 in Figure \ref{fig:model_z_plot}). In this Section, we consider the potential of detecting coherent radio emission from these events.

\subsubsection{GW170817} 
\label{sec:GW170817}
As the first confirmed binary neutron star merger, GW170817 may have emitted coherent radio emission. Although no radio telescopes were able to promptly respond to this event, we can use it to consider the coherent radio emission we would have expected. As the merger comprised of two neutron stars, we might expect to observe all of the mechanisms outlined in Section \ref{sec:priorMerge} and \ref{sec:jetInteract} depending upon the beaming properties of that emission.

Following the merger, any expected emission will rely upon the formation of a magnetar instead of a black hole. The detection of GW170817 has led to debate regarding the nature of the remnant formed by this merger \citep[e.g.][]{abbott2018}. The gravitational wave data remain inconclusive and late time electromagnetic observations do not show evidence of ongoing energy injection from the remnant \citep[e.g.][]{pooley2018}. However, the early-time optical observations of the kilonova were brighter and bluer than expected leading to suggestions that the merger remnant was a short-lived, hyper-massive neutron star \citep[e.g][]{metzger2018,ai2018}. For this system, we may therefore expect that the newly formed magnetar emitted a short-lived persistent radio pulse (as outlined in Section \ref{sec:pulsar}) until it collapsed into a black hole within a few hundred seconds after the merger. At the moment of collapse, this remnant may have also emitted a pulse of coherent radio emission (see Section \ref{sec:collapse}). 

It is possible that GW 170817 may have exhibited all of the emission mechanisms outlined in this paper, but only until a short time after the merger. Additionally, GW170817 was observed off-axis \citep{mooley2018} so, if the coherent radio emission is beamed with the same properties as the relativistic jet, it is unlikely that the radio telescopes would have detected any associated coherent radio emission even if they had responded sufficiently rapidly.

\subsubsection{Future detections}

Future gravitational wave events associated with neutron star binary mergers are going to be significantly more nearby than the cosmological GRBs discussed in Section \ref{sec:cosmoSGRBs} as the aLIGO/Virgo merger horizon is 120--170 Mpc for Observing run 3 \citep[O3;][]{abbott2018b}. Therefore, we can deduce that the emission for the models post merger are likely to be bright and detectable, even when taking into account the significant model uncertainties. However, an important issue for the gravitational wave detections is if the coherent emission is beamed. Beaming is likely for all of the emission mechanisms considered and the emission is most likely able to escape along the relativistic jet axis. For the cosmological SGRBs, we know that the system is in a favourable orientation as otherwise we would not observe the prompt gamma-ray emission. However, compact binary mergers observed via gravitational wave detections are only slightly preferred along the jet axis \citep{abadie2012} and most likely to be observed off-axis (as seen for GW170817). Thus, although the coherent radio emission is predicted to be bright for these systems, we will likely require a favourable orientation to be able to confirm or rule out the presence of this emission.

In order to constrain the prompt emission models presented in this paper, we require the rapid follow-up of SGRBs and gravitational wave-detected binary neutron stars with low frequency radio telescopes. While additional constraints can be placed on the merger remnant if the system has an X-ray light curve and known redshift, it is not known at the time of the event if such information will become available. This means that a large number of triggered follow-up observations are required to fully constrain the merger remnant X-ray and radio properties. 

Given that coherent radio emission may be emitted prior, during and after the merger, we need to trigger low frequency radio observations extremely quickly relative to the merger time and with high time resolution. SGRBs are usually at relatively high redshifts (z$\sim$0.7) so any prompt radio signals are dispersed, which causes their arrival time to be delayed by several minutes at low radio frequencies. However, the binary neutron star mergers detected by aLIGO/Virgo are far closer ($\leq$170 Mpc for O3) so any coherent signals will only be dispersion delayed by 10s of seconds. While there are several all-sky low frequency radio telescopes described in Section 3, telescopes with rapid-response triggering systems will not be able to rely on dispersion delay to provide enough time to repoint at the source. One technique that is currently being explored is to trigger follow-up radio observations on an aLIGO/Virgo negative latency alerts where gravitational waves have been detected from the binary inspiral up to 10s of seconds pre-merger. The additional time gained will allow telescopes like the MWA to trigger and be on-target to observe coherent radio emission associated with aLIGO/Virgo events \citep[see][James et al. Submitted]{chu2016}.

\section{Conclusions}

In this paper, we have considered the potential mechanisms for coherent radio emission from a compact binary merger. Many of the models continue to have large uncertainties, which are dominated by the uncertainty in the conversion of energy into observable coherent radio emission. MWA observations and X-ray modelling of GRB 150424A were used to demonstrate that the predicted emission is within reach of current rapid response radio facilities, whereas the previous limits from radio telescopes were insufficiently sensitive to constrain this emission.

By using the current knowledge of magnetar parameters from fitting the X-ray light curves of SGRBs, we determine the `typical' magnetar formed via neutron star binary mergers. This `typical' magnetar was inserted into the models, making assumptions about the efficiency of the emission mechanism, to demonstrate that the current radio facilities and with their highly complimentary observing strategies are capable of observing or placing tight constraints on the presence of this emission. For favourable orientations and sufficiently rapid response, the current radio observatories following up a gravitational wave detection of a nearby neutron star merger should be able to detect this emission or demonstrate that coherent radio emission is not able to escape from the remnant or ejecta of the merger.

\section*{Acknowledgements}
We thank the anonymous referee for their helpful comments and advice. We thank Ralph Wijers and Bradley Meyers for useful discussions. 
GEA is the recipient of an Australian Research Council Discovery Early Career Researcher Award (project number DE180100346).
This work utilises a number of Python libraries, including the Matplotlib plotting libraries \citep{hunter2007}, SciPy \citep{jones2001}, NumPy \citep{vanderwalt2011} and Sherpa \citep{freeman2001}. This work made use of data supplied by the UK Swift Science Data Centre at the University of Leicester. This research has made use of data, software and/or web tools obtained from the High Energy Astrophysics Science Archive Research Center (HEASARC), a service of the Astrophysics Science Division at NASA/GSFC and of the Smithsonian Astrophysical Observatory's High Energy Astrophysics Division.

%%%%%%%%%%%%%%%%%%%%%%%%%%%%%%%%%%%%%%%%%%%%%%%%%%



%\bibliographystyle{mnras}


%%%%%%%%%%%%%%%%%%%%%%%%%%%%%%%%%%%%%%%%%%%%%%%%%%

% Don't change these lines
\bsp	% typesetting comment
\label{lastpage}
\end{document}